 \documentclass[apj,numberedappendix]{emulateapj}

\usepackage{color}
\def\simlt{\lower.5ex\hbox{$\; \buildrel < \over \sim \;$}}
\def\simgt{\lower.5ex\hbox{$\; \buildrel > \over \sim \;$}}

\def\beq{\begin{equation}}
\def\eeq{\end{equation}}
\def\ba{\begin{eqnarray}}
\def\ea{\end{eqnarray}}

\def\qh{\dot{Q}}
\def\Fh{F_h}

\def\Sect{{\rm Section}} 
\def\Sects{{\rm Sections}} 

\def\Eq{Equation}
\def\Eqs{Equations}

\def\tauT{\tau_{\rm T}}

\def\sT{\sigma_{\rm T}}

\def\Gs{\Gamma_\star}

\def\tph{t_\star}
\def\Rph{R_\star}
\def\vsh{v_{\rm sh}}
\def\ms{m_{\star}}

\def\tobs{t_{\rm obs}}

\def\ad{\hat{\gamma}}

\def\Eb{\hat{E}}

\def\epsnu{\epsilon_{\nu}}
\def\kapnu{\kappa_{\nu}}

\def\Mej{M_{\rm ej}}

\def\vej{v_{\rm ej}}

\def\dr{\Delta r}

\def\gs{\gamma_s}
\def\vs{v_s}
\def\bs{\beta_s}

\def\mmm{m}
\def\rb{R_{\rm sh}}
\def\tej{t_{\rm ej}}
\def\mph{m_\star}

\def\tauT{\tau_{\rm T}}

\def\Epk{E_{\rm pk}}
\def\rs{r_s}

\def\brel{\beta_{\rm rel}}
\def\grel{\gamma_{\rm rel}}
\def\ggr{\chi}
\def\ms{m_s}
\def\tph{t_\star}
\def\gph{\gamma_\star}
\def\Gph{\Gamma_\star}
\def\Eph{\En_\star}
\def\Lobs{L_{\rm obs}}

\def\tdisp{t_{\rm disp}}
\def\En{{\cal E}}

\newbox\grsign \setbox\grsign=\hbox{$>$} \newdimen\grdimen \grdimen=\ht\grsign
\newbox\simlessbox \newbox\simgreatbox \newbox\simpropbox
\setbox\simgreatbox=\hbox{\raise.5ex\hbox{$>$}\llap
     {\lower.5ex\hbox{$\sim$}}}\ht1=\grdimen\dp1=0pt
\setbox\simlessbox=\hbox{\raise.5ex\hbox{$<$}\llap
     {\lower.5ex\hbox{$\sim$}}}\ht2=\grdimen\dp2=0pt
\setbox\simpropbox=\hbox{\raise.5ex\hbox{$\propto$}\llap
     {\lower.5ex\hbox{$\sim$}}}\ht2=\grdimen\dp2=0pt
\def\simgt{\mathrel{\copy\simgreatbox}}
\def\simlt{\mathrel{\copy\simlessbox}}

\begin{document}

\title{Relativistic envelopes and gamma-rays from neutron star mergers} 

\author{
Andrei M. Beloborodov,$^{1,2}$ Christoffer Lundman,$^3$ and Yuri Levin$^{1,4,5}$
}
\affil{$^1$Physics Department and Columbia Astrophysics Laboratory,
Columbia University, 538  West 120th Street New York, NY 10027 \\
% amb@phys.columbia.edu\\
$^2$Max Planck Institute for Astrophysics, Karl-Schwarzschild-Str. 1, D-85741, Garching, Germany\\
$^3$The Oskar Klein Centre, Department of Astronomy, AlbaNova, Stockholm University, SE-106 91 Stockholm, Sweden \\
$^4$Center for Computational Astrophysics, Flatiron Institute, 162 5th Ave, New York NY 10010\\
$^5$School of Physics and Astronomy, Monash University, Clayton VIC 3800, Australia
}

\begin{abstract}
We suggest that neutron star mergers eject an ultra-relativistic envelope of mass $m\sim 10^{-7}M_\odot$, which helps explain the gamma-ray burst from GW~170817. One ejection mechanism is the ablation of the neutron star surface by the burst of neutrinos in the first $30\,\mu$s of the merger. Another, more efficient, mechanism for inflating the ultra-relativistic envelope is an internal shock in the massive  ejecta from the merger. A strong shock is expected if the merger product is a magnetar, which emits a centrifugally accelerated wind. The shock propagates outward through the ejecta and accelerates in its outer layers at radii $r\sim 10^9-10^{10}$~cm, launching an ultra-relativistic opaque envelope filled with $\sim 10^4$ photons per nucleon. The Lorentz factor profile of the envelope rises outward and determines its homologous expansion, which adiabatically cools the trapped photons. Once the magnetar loses its differential rotation and collapses into a black hole, a powerful jet forms. It drives a blast wave into the envelope, chasing its outer layers and eventually catching up with the envelope photosphere at $r\sim 10^{12}$~cm. The ultra-relativistic photospheric breakout of the delayed blast wave emits a gamma-ray burst in a broad solid angle around the merger axis. This model explains the gamma-ray pulse from merger GW~170817 with luminosity $L_\gamma\sim 10^{47}$~erg/s, duration $\Delta \tobs\sim 0.5$~s, and characteristic photon energy $\sim 100$~keV. The blast wave Lorentz factor at the envelope photosphere is consistent with $\Gamma\simgt 5$ that we derive from the observed light curve of the burst. We suggest future tests of the model.
\end{abstract}


%#######################################################################

\section{Introduction}

\subsection{Ejecta from neutron star merger GW~170817}

The recent detection of gravitational waves from neutron star merger GW 170817 and its electromagnetic counterpart opens a new window for the studies of neutron stars, cosmological gamma-ray bursts (GRBs), and the origin of heavy nuclei \citep{Abbott2017a,Abbott2017b,Goldstein2017,Savchenko2017,Coulter2017,Evans2017,Soares-Santos2017}. The electromagnetic radiation was emitted by ejecta from the merger, viewed at an angle of $\theta\sim 20^\circ-30^\circ$ from the rotation axis of the binary. The viewing angle and the distance to the merger $d\sim 40$~Mpc are both estimated from the observed gravitational wave signal, and its host galaxy was found at $d\approx 40$~Mpc.

The gamma-ray counterpart, GRB~170817A, had luminosity $L\sim 10^{47}$~erg/s and was emitted with a delay of $\sim 1.7$~s following the merger. It could be powered by a delayed jet from the central remnant when it breaks out from a massive cloud around the remnant \citep{Kasliwal2017,Gottlieb2018,Bromberg2018, Pozanenko2018}.

A day later, the cloud of expanding ejecta with mass $\Mej\sim 5\times 10^{-2}M_\odot$ and speed $v/c\sim 0.1-0.3$ emitted optical radiation with luminosity $L\sim 10^{41}$~erg/s --- the ``kilonova.'' Its light curve was consistent with being powered by the decay of r-process nuclei \citep{Kasen2017,Drout2017,Tanaka2017,Tanvir2017}.
 
 At yet later times (weeks), X-ray and radio afterglow was observed \citep{Troja2017,Margutti2017,Hallinan2017}. The unusually late rising afterglow was proposed to result from deceleration of quasi-isotropic, moderately relativistic ejecta from the merger in an external medium \citep{Nakar2018b}. It was also found consistent with a decelerating ultra-relativistic narrow jet with energy $E_{\rm jet}\sim 10^{50}$~erg  \citep{Lazzati2018,Granot2018,Lamb2018,Xie2018}, and further evidence for a collimated jet came from radio imaging \citep{Mooley2018a,Mooley2018b}. The jet is initially invisible to off-axis observers, because of its strong collimation and Doppler-beaming with a very high Lorentz factor. Its emission comes into view after significant deceleration, long after the merger.

\subsection{The puzzling GRB~170817A}

Generally, emission from GRB jets was expected to be weak and soft when viewed off-axis (e.g. \citealt{Lazzati2017a,Lazzati2017b}). By contrast, the gamma-ray pulse of GRB~170817A is not soft; its spectrum peaks at $100-200$~keV (Goldstein et al. 2017), and an even harder spectrum was detected in a short time interval of $\simlt 0.1$~s \citep{Veres2018}. Furthermore, GRB jets display a strong correlation between the apparent brightness and hardness of their emission; GRB~170817A is far off this correlation (\citealt{Pozanenko2018}).

In addition, the simple light curve of GRB~170817A favors a single emission event, such as a blast wave from a jet, rather than variable internal dissipation typical for GRB jets. The canonical GRBs viewed on-axis are extremely bright and have diverse (usually multi-peaked) light curves \citep{Nakar2007}. The off-axis GRB~170817A is dominated by a single weak pulse of width $\Delta \tobs\sim 0.5$~s, smaller than but comparable with its delay $\tobs\sim 1.7$~s. Goldstein et al. (2017) also reported an unusual transition from the gamma-ray pulse to a  quasi-blackbody X-ray tail, although the tail has a low signal-to-noise ratio and its detailed spectral shape is uncertain.

A natural mechanism for a single, hard, gamma-ray pulse followed by a soft thermal tail could be the breakout of a shock wave from the massive cloud around the merger. Following previous theoretical calculations (e.g. \citealt{Nakar2012}), the shock breakout model proposed for GRB~170817A \citep{Kasliwal2017,Gottlieb2018,Bromberg2018} posits the shock temperature $kT\approx 50$~keV. It predicts the observed (Doppler-shifted) spectral peak at $3kT\Gamma\sim 150$~keV if the Lorentz factor of the shock-heated plasma is  $\Gamma\sim 1$. The low $\Gamma$ is, however, in conflict with observations. In \Sect~2 we show that the light curve of GRB~170817A requires the gamma-ray source to have $\Gamma\simgt 5$.

The Lorentz factor $\Gamma\sim 5$ could be consistent with an off-axis  component (``cocoon'') of an ultra-relativistic jet after its breakout from the massive cloud. However, this outflow is not expected to emit gamma-rays. Simulations of jet breakout and cocoon expansion show that the off-axis outflow is heated by internal shocks too early, before it becomes transparent to radiation, and this leads to reprocessed X-ray emission rather than gamma-ray emission \citep{Lazzati2017a,Lazzati2017b}.

\subsection{This paper}

After estimating the lower limit on $\Gamma$ in GRB~170817A in \Sect~2, we turn to the theory of relativistic ejecta from neutron star mergers. We find that, before jets are launched and  the GRB is emitted, the merger is likely to eject an ultra-relativistic opaque envelope. The envelope quickly expands around from the central massive cloud and thus greatly inflates the photospheric radius of the merger ejecta. In \Sects~3 and 4  we describe two mechanisms for the envelope ejection, and calculate its expected self-similar  structure. In both cases, we find a stratified structure with four-velocity $\gamma\beta$ growing outward and extending to $\gamma\beta\gg 1$. We estimate the expected mass of the ultra-relativistic envelope and its photon-to-baryon ratio. 

The first mechanism of the envelope ejection is the ablation of neutron star surface at the very beginning of the merger, when it suddenly (in $30\,\mu{\rm s}$)
becomes a powerful source of neutrinos (\Sect~3). The reaction  $\nu\bar{\nu}\rightarrow e^\pm$ injects heat near the surface of the merging stars, resulting in huge energy per baryon and accelerating the surface layers to ultra-relativistic speeds. This effect is missed by the existing merger simulations (e.g. \citealt{Dessart2009,Bauswein2013,Hotokezaka2013,Hotokezaka2018,Radice2016,Radice2018,Kiuchi2018}),
 because they do not resolve the heating and dynamics of low-density surface layers. We find that an ultra-relativistic ablated mass $m\simgt 10^{-8}M_\odot$ promptly escapes the vicinity of the merger, before the outflow becomes heavily polluted by baryons forming the massive cloud of slower ejecta.

The second mechanism is described in \Sect~4. The expanding cloud of massive ejecta can develop a strong internal shock, which propagates outward and accelerates to ultra-relativistic speeds in the outermost, low-density layers of the cloud. One appealing scenario invokes the formation of a rapidly spinning, short-lived magnetar following the merger. The fast outflow from the magnetar drives a shock into the cloud, which appears favorable  for production of the ``blue'' kilonova \citep{Metzger2018}. We show that after the shock crosses the cloud and accelerates in its outer layers, an ultra-relativistic envelope is inflated. We estimate its mass $m\simgt 10^{-7}M_\odot$ and describe its self-similar structure.

The presence of the envelope weakly affects the bright beamed GRBs seen by on-axis observers, however it can play a key role in off-axis GRB production. In \Sect~5 we discuss how a delayed launch of a jet inside the envelope leads to the production of an off-axis, single-pulse GRB, and compare our model predictions with GRB~170817A. We find that the blast wave from the jet in the envelope can explain the observed luminosity, hardness, and light curve of the burst. Comparison with previous models and observational implications are discussed in \Sect~6.

%#######################################################################

\section{Relativistic motion in GRB~170817A}

\cite{Kasliwal2017} found a lower limit for the Lorentz factor of the GRB source $\Gamma>2.5$. They used the considerations of photon-photon opacity due to collisions $\gamma\gamma\rightarrow e^+e^-$, which occur for photons of energies above $E_{\rm thr}\sim \Gamma$~MeV (e.g. \citealt{Lithwick2001}). It is, however, difficult to derive robust limits from photon-photon collisions. In principle, the source is allowed to be completely opaque to MeV photons, as no such photons were observed, and their number can only be guessed from extrapolations of the observed spectrum.

Instead, a simple lower bound on $\Gamma$ can be obtained from the standard consideration of the scattering opacity in the source (e.g. \citealt{Lithwick2001}). Let $L\equiv 4\pi r^2 \Gamma^2 n_b m_pc^3$ be the isotropic equivalent of the kinetic power of the relativistic outflow emitting the observed gamma-rays with luminosity $L_\gamma$ (note that $L$ does not include $L_\gamma$ and in general may be smaller or larger than $L_\gamma$). Here $r$ is the emission radius, $n_b$ is the baryon number density in the flow rest frame, and $m_p$ is the proton mass. A characteristic optical depth to Thomson scattering is given by
\beq
\label{eq:tauT}
   \tauT=\frac{\zeta\sT n_p r}{\Gamma}=\frac{\zeta \sT L Y_e}{4\pi r m_pc^3\Gamma^3},
\eeq
where $n_p=Y_e n_b$ is the proton density, $Y_e$ is the proton-to-nucleon ratio, and $\zeta\simlt 1$ is a numerical coefficient, which depends on the radial density profile of the outflow. 

If $L$ is approximately uniform on radial scales $\Delta r>r/\Gamma^2$ then one can show that $\zeta\sim 1$. This situation is expected if the outflow is launched on a timescale $\Delta t > r/\Gamma^2 c$. In particular, a ballistic outflow with $\Gamma\approx const$ and $L\approx const$ has a density profile $\rho\propto r^{-2}$. The characteristic optical depth for photons (emitted isotropically in the fluid frame) can be found by integrating the scattering coefficient $\alpha_{\rm sc}\propto\rho$ along the photon trajectory. This calculation gives \Eq~(\ref{eq:tauT}) with $\zeta\sim 1$ \citep{Abramowicz1991,Beloborodov2011}. 

In the opposite limit, one can consider an outflow ejected impulsively, within $\Delta t\approx 0$. Then its density profile is set by radial spreading during the outflow expansion to the radius of GRB emission. This radial spreading is controlled by the distribution of four-velocity, which has a significant width for any realistic ejection mechanism. In particular, the homologous envelope described later in this paper can be idealized as an impulsive ballistic ejection with a power-law distribution of four-velocity (see \Sect~5). A photon propagating in the homologous envelope will see a steeply decreasing density. In this case $\zeta$ can be as low as $1/4$.

The GRB radiation can escape if $\tauT\simlt 1$, which gives a lower limit on $\Gamma$,
\beq
\label{eq:Gs}
   \Gamma>\Gs\approx 4.9\,r_{12}^{-1/3} \left(\frac{\zeta Y_eL}{L_\gamma}\right)^{1/3} L_{\gamma,47}^{1/3}.
\eeq

%%%%%%%%%%% FIGURE %%%%%%%%%%%%%%%%%%
\begin{figure}[t]
\includegraphics[width=0.47\textwidth]{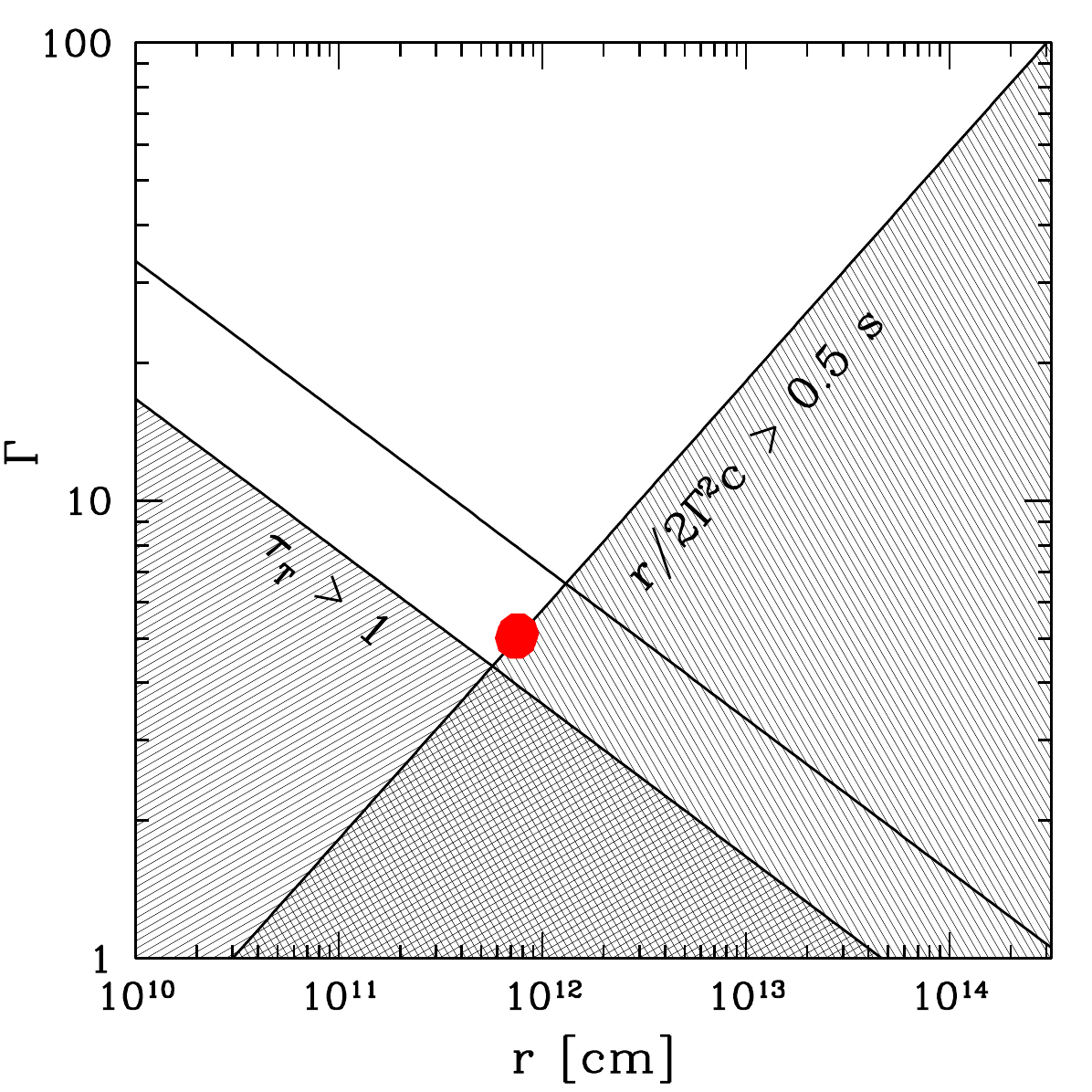}
\caption{Constraints on the radius $r$ and Lorentz factor $\Gamma$ of the plasma emitting GRB~170817A. The shaded regions are excluded by the observed peak duration $\Delta t_{\rm obs}\sim 0.5$~s and the optical depth requirement $\tauT\simlt 1$. The line of $\tauT=1$ is shown for two cases: $\zeta Y_e L/L_\gamma =2$ (upper line) and 1/4 (lower line). Red circle shows the parameters of the burst model described in Section~5.
}
\label{ifig:constraints}
 \end{figure}
%%%%%%%%%%% FIGURE %%%%%%%%%%%%%%%%%%

Another constraint is set by the minimum dispersion of the photon arrival times $\Delta\tdisp\sim r/2\Gamma^2 c$. It applies as long as the GRB-emitting shell has a minimum angular size $\Delta \theta_{\min}\sim \Gamma^{-1}$, which is valid for any expanding relativistic cloud accelerated by its internal pressure.\footnote{The angular size of the emitting shell in GRB~170817A is likely to significantly exceed $\Gamma^{-1}$ (below we argue that $\Gamma\simgt 5$). The fact that we receive radiation at the polar angle $\theta\sim 0.5$~rad, as inferred from the gravitational wave signal, suggests $\Delta\theta\simgt 0.5>\Gamma^{-1}$.} 
The minimum dispersion $\Delta \tdisp$ should not exceed the observed duration of the main peak of GRB~170817A $\Delta \tobs\sim 0.5$~s, which gives the constraint
\beq
\label{eq:dt}
   r\simlt 2\Gamma^2 c\,\Delta \tobs 
     = 3\times 10^{10}\,\Gamma^2 \left(\frac{\Delta \tobs}{0.5\rm~s}\right) {\rm~cm}.
\eeq
Note that only the duration of the peak is relevant for this constraint; it is not affected by the arrival time of the peak $\tobs\sim 1.7$~s.

Combining the two constraints in \Eqs~(\ref{eq:Gs}) and (\ref{eq:dt}), and using the observed $L_\gamma\approx 1.6\times 10^{47}$~erg/s and $\Delta\tobs\approx 0.5$~s \citep{Goldstein2017}, we find
\beq
  \Gamma>5.7
  \left(\frac{L_\gamma}{1.6\times 10^{47} \rm erg/s}\right)^{1/5}
  \left(\frac{\Delta \tobs}{0.5\rm~s}\right)^{-1/5}
  \left(\frac{\zeta Y_e L}{L_\gamma}\right)^{1/5}.
\eeq 

Figure~1 shows the constraints on the GRB source in the $r$-$\Gamma$ plane. If most of the observed luminosity comes from the photosphere of the explosion, the source must be located near the line of $\tauT=1$. If it is produced by a photospheric shock breakout then it should also be located near the line of $r/2\Gamma^2 c=0.5$~s, and so in this case it will be near the intersection of the two lines. 

A similar minimum $\Gamma\sim 5$ for GRB~170817A may be obtained by considering the scattering opacity of $e^\pm$ pairs created by MeV photons, using a high-energy extrapolation of the observed spectrum \citep{Matsumoto2019}.
 
One could also consider the possibility that the source is magnetically dominated, i.e. powered by Poynting flux from the central engine and carries a negligible amount of baryonic matter. Then the plasma emitting the GRB would be made entirely of $e^\pm$ pairs created in photon-photon collisions. The source becomes transparent and releases radiation when its compactness parameter $\ell=L_\gamma\sT/m_ec^3 R\Gamma^3$ decreases to $\sim 10$,
which requires $\Gamma>10\, r_{11}^{-1/3}$.

%#######################################################################

\section{Ablation of the neutron star surface}

\subsection{Hot sandwich at the collision interface}

The two merging stars are strongly deformed by tidal forces, forming cusps pointing approximately toward each other (but still significantly misaligned, because of the orbital rotation of the binary). The merger begins with the tangential collision of the cusps, forming a growing interface between them (see e.g. Figure~4 in \citealt{Bauswein2013}). The surface layers at the interface are shocked, compressed, and heated, forming a thin ``sandwich.'' Local thermodynamic equilibrium and nuclear statistical equilibrium are quickly established in the sandwich, with pressure contributions from nuclei, electrons, positrons, and Planckian radiation, at a high temperature $T$. 

The opposite tangential velocities $v_\parallel$ of the colliding stars imply a huge velocity shear at the interface, which immediately leads to Kelvin-Helmholtz instability \citep{Price2006,Kiuchi2018}, with a growth rate comparable to the shear rate. The limited numerical resolution of the global merger simulations makes it difficult to observe the fast shear damping that develops on smallest scales, and local simulations \citep{Zrake2013} show more details of the Kelvin-Helmholtz instability. The efficient damping of the tangential motion suggests that the collision is ``sticky,'' releasing its entire specific energy $v_\star^2/2=(v_\parallel^2+v_\perp^2)/2$, not just the normal component $v_\perp^2/2$. Then the energy released in the sandwich per unit baryon rest mass is given by
\beq
\label{eq:Eb}
  \Eb\approx \frac{v_\star^2}{2c^2}. 
\eeq
It may exceed 0.03 in mergers of massive neutron stars.

The hot compressed sandwich is bounded by two shocks propagating into the stars with speed $\vsh\sim v_\star$. The energy density in the sandwich is $U\approx \xi\rho c^2 \Eb$, where $\rho$ is the upstream (pre-shock) density, $\xi=(\ad+1)/(\ad-1)$ is the shock compression factor, and $\ad$ is the adiabatic index of the post-shock matter. The sandwich pressure is 
\begin{eqnarray}
\nonumber
   P &=& (\ad-1) U\approx (\ad+1)\rho c^2 \Eb \\
      &\approx & 8\times 10^{29}\,\rho_{10}\left(\frac{\Eb}{0.03}\right) {\rm erg~cm}^{-3}.
   \label{eq:P}
\end{eqnarray}
As the two shocks propagate into denser subsurface layers of the colliding stars, the sandwich pressure grows, $P\propto\rho$. 

The approximate pressure balance between the shocks implies $P(x)\approx const$ across the sandwich, where the $x$-axis is chosen normal to the collision interface.\footnote{Variation of $P$ in the $x$-direction is small because the shocks are in hydrodynamical causal contact and not far from pressure equilibrium. At the same time, $P(y,z)$ strongly varies along the collision interface, decreasing from the center (the initial touch point of the colliding stars) to the outer parts of the sandwich, where shocks form later (Figure~2). Initially the collision interface area grows superluminally, faster than the shocked matter could be squeezed out from the sandwich. At a later stage, the pressure gradient in the $y$-$z$ plane begins to drive a fan-like ``fountain'' from the sandwich.} However, the nucleon density is strongly non-uniform in the $x$ direction --- the sandwich is made of layers of stratified density. The older layers in the middle were shocked at a low pressure and later pressurized through compression. As $P$ grows proportionally to the upstream density $\rho$, the old shocked layers are strongly compressed to stay in the approximate pressure equilibrium with the propagating shocks. This compression implies strong adiabatic heating, which produces a low-mass, thin layer of ultra-relativistic material. The compressional heating is discussed in more detail in Appendix.

In principle, the compressed layer heated to specific enthalpy $W>c^2$ could be partially ejected with an ultra-relativistic speed. The basic effect may be illustrated by a vessel of hot gas compressed to a small volume by external pressure. The work performed to compress the vessel is stored in the gas internal energy, and a sufficiently strong compression makes the gas relativistically hot,  $W\gg c^2$. When the external pressure is eventually removed, the gas will explode and achieve ultra-relativistic speeds. This toy model does not, however, capture additional effects expected at the interface of the colliding stars. In particular, mixing and transport effects should suppress the relativistic ejection (see discussion in Appendix).

The magnetospheres of the colliding neutron stars will also be strongly compressed in collision. In an idealized model, this would create a magnetic ``pillow'' at the interface between the stars, with magnetic pressure in an approximate balance with the ram pressure of the two shocks propagating into the stars $B^2/8\pi\sim P$. As the ram pressure grows up to $P\sim 10^{34}\rho_{14}$~erg~cm$^{-3}$ (see \Eq~\ref{eq:P}), the magnetic field in the pillow is amplified up to $B\sim 5\times 10^{17}\,\rho_{14}^{1/2}$~G, where $\rho$ is the matter density upstream of the shocks. This implies compression of the magnetosphere by a huge factor $B/B_0$ for reasonable pre-merger magnetic fields $B_0$. The resulting pillow thickness $\delta\sim (B_0/B)R$ is many orders of magnitude smaller than the stellar radius $R\sim 10$~km.

However, the idealized picture of a compressed magnetic pillow is destroyed by instabilities of the Raleigh-Taylor type, which will tend to mix the compressed magnetic field into dense stellar material. Note also that the magnetic field is amplified in a much thicker layer as a result of Kelvin-Helmholtz instability.

\subsection{Neutrino emission from the sandwich}

When the two stars have just touched, the sandwich pressure and temperature are initially modest; at this earliest stage most of the collision energy converts to Planckian radiation. When the shocks propagate into deep and dense layers, $\rho\simgt 10^{12}$~g~cm$^{-3}$, the pressure becomes dominated by nucleons rather than radiation. The simplest estimate for the sandwich temperature is given by the upper limit,
\beq
  kT_{\max}\equiv \frac{2}{3}\,\Eb\, m_n c^2\approx 20\,\left(\frac{\Eb}{0.03}\right){\rm~MeV} 
  \quad {\rm (no~cooling)},
\eeq 
which neglects any contributions to pressure other than nucleons. 

\begin{figure}[t]
% \begin{center}
\includegraphics[width=0.47\textwidth]{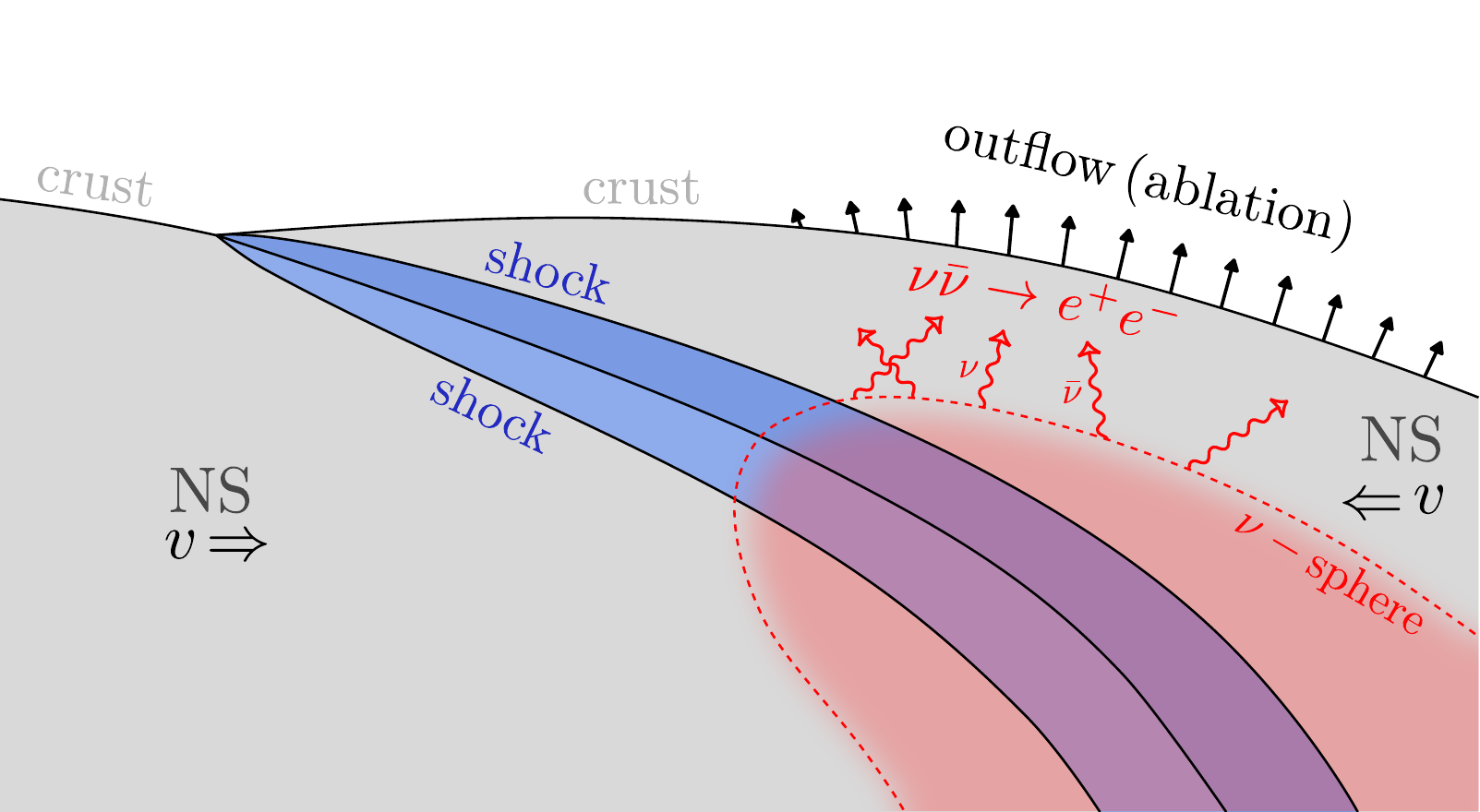} 
% \end{center}
\caption{Schematic picture of the shocked sandwich at the collision interface (blue) and its neutrino emission (red). The neutrinos decouple from the stellar matter (shaded in gray) and escape with a quasi-thermal spectrum at the neutrino-sphere (dotted red curve). Some of the emitted $\nu$ and $\bar{\nu}$ collide outside the neutrino-sphere and convert to $e^\pm$ pairs. This process deposits energy into the cold surface layers and leads to their ablation.}
\label{scheme}
 \end{figure}

The rate of neutrino and antineutrino production by the sandwich is quickly increasing with temperature $T$, and becomes significant before $T$ approaches $T_{\max}$. The cooling becomes significant when the shocks reach crustal layers with densities $\rho\gg 10^{10}$~g~cm$^{-3}$; then the cooling timescale becomes  shorter than the shock age (the details will be described elsewhere). Neutrinos are mainly produced by the $e^\pm$ capture reactions $e^++n\rightarrow p+\bar{\nu}$ and $e^-+p\rightarrow n+\nu$. If the neutrinos escape, their energy flux can be estimated as 
\beq
\label{eq:F}
   F\sim \frac{\rho v_\star^3}{2}  {\rm~~~~(efficient~cooling)}.
\eeq
This simply states that most of the energy released in the shock converts to the neutrino flux.  
The effective temperature $T$ of escaping neutrinos is approximately related to their energy flux by $F\approx \sigma T^4$, where $\sigma=ac/4$ is the Stefan-Boltzmann constant.\footnote{The emitted neutrino flux is not exactly $\sigma T^4$ for two reasons: (i) their spectrum is not exactly thermal, and (ii) even for completely thermalized neutrinos, their (fermion) statistics is different from photon statistics. The numerical factor resulting from these corrections is $\sim 1$ and weakly affects the estimate in \Eq~(\ref{eq:Th}).}
This gives an estimate,
\begin{equation}
\label{eq:Th}
  kT\approx 15\,\rho_{12}^{1/4}\,v_{\star,10}^{3/4}{\rm ~MeV} 
   {\rm~~~~(efficient~cooling)}.
\end{equation}
It may be viewed as a lower bound on $T$ at large densities. The upstream $\rho$ appearing in \Eqs~(\ref{eq:F}) and (\ref{eq:Th}) is lower near the edges of the sandwich, where the shocks formed later and therefore had less time to propagate into deep subsurface layers (Figure~\ref{scheme}). The sandwich size measured along the collision interface  grows from the initial contact point to $\sim 10$~km on a timescale $t\sim 3\times 10^{-5}$~s.

%%%%%%%%%%% FIGURE %%%%%%%%%%%%%%%%%%
%%%%%%%%%%% FIGURE %%%%%%%%%%%%%%%%%%

As the two shocks bounding the sandwich propagate into the subsurface layers of increasing density $\rho$, the post-shock temperature grows and so do the energies of emitted neutrinos $\epsnu$ (in units of $m_ec^2$). The cross section for neutrino interaction with matter grows as $\epsnu^2$, and the neutrinos eventually become absorbed near (or inside) the sandwich. This occurs when the shocks propagate sufficiently deep below the stellar surface, $\rho\gg 10^{11}$~g~cm$^{-3}$. Then neutrino transport occurs in a diffusive regime, with the local neutrino density $U_\nu$ close to local thermodynamic equilibrium. The mean energy of the thermalized neutrinos (in units of $m_ec^2$) is
\beq
   \epsnu\approx 3\,\Theta\equiv 3\,\frac{kT}{m_ec^2} {\rm~~~~(thermalized).}
\eeq

Cross sections for neutrino interactions with nucleons and leptons are summarized e.g. in \cite{Chen_Beloborodov2007}. For our estimates it is sufficient to include one process --- neutrino absorption by nucleons, which has the largest opacity $\kapnu\approx 3\times 10^{-20}\,\Theta^2$~cm$^2$~g$^{-1}$. The mean free path of neutrinos is 
\beq
\label{eq:lnu1}
  l_\nu=\frac{1}{\kappa_\nu\rho}
       \approx 3.6\times 10^5\,\rho_{11}^{-1}\left(\frac{\Theta}{30}\right)^{-2}{\rm cm}. 
\eeq
One can see that at densities $\rho\gg 10^{11}$~g~cm$^{-3}$ the neutrinos are self-absorbed and thermalized.

At the advanced stage of collision, most of the produced neutrinos are trapped in the middle of the sandwich, and the escaping neutrinos diffuse out from its outer parts (Figure~\ref{scheme}), where density $\rho\simlt 10^{12}$~g~cm$^{-3}$. The simple estimate in Equation~(\ref{eq:F}) then suggests a characteristic value for the escaping flux $F\sim 10^{41}$~erg~s$^{-1}$~cm$^{-2}$. This estimate is, however, crude. Accurate spatial distribution of $F$ could be found with expensive three-dimensional simulations of the collision dynamics and neutrino transport. The result will depend on the masses of the colliding stars. Massive mergers produce high $F$ because they have a high collision speed $v_\star$.

The flux $F\sim 10^{41}$~erg~s$^{-1}$~cm$^{-2}$ is emitted from the ``neutrino-sphere'' --- the surface where the neutrinos decouple from matter  and begin to stream freely. The temperature of the neutrino-sphere $T_\star$ satisfies the approximate relation $F\sim \sigma T_\star^4$ which gives $\Theta_\star\sim 30$.

%########################################################

\subsection{Heating of surface layers by reaction $\nu\bar{\nu}\rightarrow e^+e^-$}

Some of the neutrinos and anti-neutrinos escaping from the neutrino-sphere can collide with each other and convert to $e^\pm$ pairs, depositing heat. The rate of this ``neutrino heating'' is independent of the local matter density and in the low-density regions near the stellar surface it injects huge energy per nucleon, rising the local specific enthalpy,
\beq
  w\equiv \frac{U+P}{\rho c^2}
\eeq
to relativistic values $w\simgt 1$. The heated surface layers will expand with relativistic speeds and leave the star. Thus a fraction of the stellar crust will be ablated by neutrino heating. 

Neutrinos collide and turn into $e^\pm$ pairs with a significant cross section when there is a significant angle between their directions, $\delta\simgt 1$. The neutrino heating wave propagates with a speed $v\sim c \cos\delta$, comparable to $c/2$. The wave is faster than the shocks, so ablation of surface layers occurs before the shock arrival (Figure~\ref{scheme}).

We wish to obtain a rough estimate for the mass ablated to highly relativistic speeds. The first step is to evaluate the heating rate $\qh$ due to reaction $\nu\bar{\nu}\rightarrow e^+e^-$. The cross section for this reaction  is given by \citep{Goodman1987},
\beq
  \sigma_{\nu\bar{\nu}}\approx 0.2\,\sigma_0\,
  \frac{ c^4({\mathbf p}_{\nu}\cdot {\mathbf p}_{\bar\nu})^2}{(E_\nu E_{\bar\nu})^2}
  =0.2\,\sigma_0\,\epsnu\epsilon_{\bar\nu}\,(1-\cos\delta)^2,
\eeq 
where $\sigma_0\approx 1.7\times 10^{-44}$~cm$^2$, $\epsilon=E/m_ec^2$ (with subscripts $\nu$ and $\bar\nu$ corresponding to the colliding neutrino and anti-neutrino), ${\mathbf p}=(E/c,p_x,p_y,p_z)$ is the four-momentum of the colliding neutrino/anti-neutrino, and $\delta$ is the angle between their directions. 

Each reaction $\nu\bar{\nu}\rightarrow e^+e^-$ creates an $e^\pm$ pair with total energy $E_\nu+E_{\bar\nu}$. The corresponding energy deposition rate per unit volume $\qh$ may be estimated by replacing the quasi-thermal spectrum of the neutrino-sphere by a delta-function at the average neutrino energy, $\epsnu\approx\epsilon_{\bar\nu}\approx 3\Theta_\star$. Then one finds
\beq
\label{eq:qh1}
  \qh\sim \sigma_{\nu\bar{\nu}}c\,n_\nu n_{\bar\nu}(E_\nu+E_{\bar\nu})
  \sim \frac{\sigma_{\nu\bar{\nu}}}{c}\, \frac{F_{\nu}}{E_\nu}\,
  \frac{F_{\bar\nu}}{E_{\bar\nu}}(E_\nu+E_{\bar\nu}).
\eeq
We will also use the estimate $F_\nu\approx F_{\bar\nu}=F\approx \sigma T_\star^4$. This gives
\beq
\label{eq:qh}
   \qh\sim \sigma_0\overline{(1-\cos\delta)^2}\,\frac{\Theta_\star F^2}{m_ec^3}
     \sim 10^{35} F_{41}^{9/4} {\rm erg~s}^{-1}{\rm cm}^{-3},
\eeq
where $\overline{(1-\cos\delta)^2}\simlt 1$ is a numerical factor obtained after averaging over the directions of the colliding neutrinos. 

It may be instructive to compare the estimate~(\ref{eq:qh}) with previous numerical simulations of {\it steady} heating by the reaction $\nu\bar{\nu}\rightarrow e^+e^-$ by \cite{Birkl2007} and \cite{Zalamea2011}, who focused on accretion disks around accreting black holes, in Kerr metric. \cite{Birkl2007} also calculated steady heating in spherical geometry, with and without gravitational bending of neutrino trajectories. These simulations gave the efficiency of converting neutrino flux to heat, $\Fh/F$, where 
\beq
  \Fh=\int \qh\,ds\sim H\qh,
\eeq
$s>0$ is the altitude above the neutrino source, $\Fh$ is the vertically integrated heating rate, and $H$ is the characteristic scale-height of the heating region. In the simplest spherical model, the thermal neutrino source is described by its surface flux $F$ and radius $R$. The characteristic $H$ is comparable to $0.1R$. We have checked that the heating rate calculated by Birkl et al. (2007), scaled to $R\sim 10$~km and $F\sim 10^{41}$~erg~s$^{-1}$~cm$^{-2}$), is approximately consistent with the estimate in \Eq~(\ref{eq:qh}). It gives a rough estimate of the heating efficiency $F_h/F\sim H\qh/F$ approaching 0.1.
 
In a real merger the neutrino source geometry is neither spherical nor axisymmetric, and the neutrino-sphere is not parallel to the stellar surface (Figure~\ref{scheme}). Furthermore, an essential difference from the previous work is that here we deal with an initial-value problem rather than a steady state. The ablation of surface layers is triggered by the suddenly arising burst of neutrinos from the sandwich. 

It takes a very short time $t\sim x/\vsh\sim 10^{-5}$~s for the sandwich density and temperature to reach high values so that its neutrino flux approaches $F\sim 10^{41}$~erg~s$^{-1}$~cm$^{-2}$. As the sandwich size grows, the area of the neutrino-sphere grows on a similar timescale. Thus, the local $\qh(t)$ measured at the stellar surface is a steeply increasing function of time, shaped by the evolution of the neutrino-sphere and the emitted flux $F$. This function is also slightly affected by a propagation delay: the wave of heat injection propagates with a speed of $v_h\sim c\cos\delta\simlt c$ from the neutrino-sphere to the stellar surface.

\subsection{Estimate for relativistically ablated mass}

Only the uppermost ablated layers, which have low densities, reach highly relativistic speeds. The relativistic ablation ends after a short time $t_a$ when sufficient amount of matter is lifted from the NS surface and fills the main heating region $H\sim 1$~km. Later the neutrino-driven outflow becomes a relatively slow quasi-steady wind, which was studied previously in detail \citep{Qian1996,Thompson2001,Thompson2004, Metzger2007,Dessart2009}. The quasi-steady wind at $t\gg t_a$ is mainly heated through neutrino absorption by baryons. By contrast, heating at the initial ablation stage $t<t_a$ is dominated by the process $\nu\bar{\nu}\rightarrow e^+e^-$. There is much less matter in the initial relativistic outflow, and so neutrinos are mainly absorbed in $\nu\bar{\nu}$ collisions rather than by baryons.

Ablation may be roughly described as a two-step process: (1) enthalpy $w$ is deposited in the heating zone $s\simlt H\sim 1$~km, where matter begins its acceleration, and (2) adiabatic expansion at $s>H$ converts enthalpy to bulk kinetic energy. For instance, layers with asymptotic $\gamma\beta\sim 1$ are still relatively slow in the heating zone $s\simlt H$. Their modest characteristic speed $v_a\sim (0.1-0.2)c$ gives them time $t_a\sim H/v_a\simgt 10^{-5}$~s to accumulate enthalpy $w\sim 1$ before leaving the heating zone.

One can estimate the mass of ultra-fast ablated layers from the condition $w\simgt 1$ in the heating zone. The deposited energy density 
\beq
   U\sim\qh t_a\sim 10^{30}\,F_{41}^{9/4}t_{-5} {\rm~erg~cm}^{-3}
\eeq 
will give dimensionless enthalpy $w>1$ in layers of density $\rho<(U+P)/c^2\sim 10^9$~g~cm$^{-3}$. These layers occupy an initial volume $V\sim A h\simgt 10^{16}$~cm$^3$, where $A\sim 10^{11}-10^{12}$~cm$^2$ is the area of the neutrino-sphere near the sandwich edge, where ablation occurs, and $h\sim 0.3-1$~km is the characteristic hydrostatic scale-height of the stratified crust in the colliding tidal cusps of the stars. This gives the relativistically ablated mass $\sim V\,\qh t_a/c^2$.

The relativistic ablation is not isotropic and may peak in a solid angle $\Omega_a\sim 1$, which corresponds to a beaming factor of $b\sim 4\pi/\Omega_a$. The isotropic equivalent of the ablated mass $m$ viewed within $\Omega_a$ is given by
\beq
\label{eq:m}
  m\sim b\,V\,\frac{\qh t_a}{c^2}\sim 10^{-7}\,M_\odot\, b_1 V_{16}\,F_{41}^{9/4}. 
\eeq
We conclude that the observed isotropic equivalent of the ablated mass with $\gamma\beta> 1$ may exceed $10^{26}$~g, depending on the viewing angle and the precise values of the beaming factor $b$ and the neutrino-sphere flux $F$.

\subsection{Numerical simulation}

The features of relativistic ablation discussed above are illustrated by the following simplified numerical model. Let us replace the merging stars by a single sphere of radius $R$ and choose the surface heating rate $\qh$ in the form
\beq
\label{eq:qh1}
  \qh(t,s)=\qh_0(t) \left(\frac{a+s}{a}\right)^{-4}\left[1-\frac{s}{(s^2+a^2)^{1/2}}\right]^2.
\eeq
Here $s=r-R_\star>0$ is the distance from the neutrino-sphere $R_\star$, which is somewhat below the stellar surface, $R_\star<R$. Our simulation assumes the neutrino-sphere radius $R_\star=10$~km and the star radius $R=11$~km.

A true spherical source of neutrinos would have $a=R_\star$ in \Eq~(\ref{eq:qh1}). However, our spherical model is designed as a proxy for the colliding stars, and we use $a<R_\star$ to parameterize the relatively small size of the ablation region near the sandwich edge (Figure~\ref{scheme}). The dependence of $\qh$ on $s$ has two parts: the power law $(1+s/a)^{-4}$ describes the reduction of $n_\nu n_{\bar\nu}$ with distance from the neutrino-sphere, and the term in square brackets roughly describes the dependence $\sigma_{\nu\bar\nu}\propto \overline{(1-\cos\delta)^2}$, where $\tan\delta\sim a/s$ represents the characteristic angle between the colliding neutrinos. \Eq~(\ref{eq:qh1}) implies a steep decline of $\qh$ with $s$. This leads to a modest characteristic thickness of the main heating region, comparable to or smaller than 1~km. 

The $\qh_0(t)$ in \Eq~(\ref{eq:qh1}) describes the time dependence of surface heating by the neutrino wave from the sandwich. At early times, $\qh_0(t)$ is a steep function, because it is proportional to $F^{9/4}$, and $F$ is quickly increasing as the sandwich pressure grows, $P\propto \rho$ (\Eq~\ref{eq:P}).  We do not know the exact shape of this function and replace it with a simple power law,
\beq
  \qh_0(t)=10^{35}\left(\frac{t}{t_0}\right)^\zeta {\rm~erg~s}^{-1}{\rm~cm}^{-3}, \qquad t<t_0,
\eeq
followed by constant $\qh_0(t)=\qh_0(t_0)$ at $t>t_0$. Our sample model will have $\zeta=9$ and $t_0=10^{-5}$~s. This crude description of the heating onset captures its main feature: the steep rise on a short timescale, which will lead to relativistic ablation of the surface layers.

The simulation tracks the dynamics of the outer crust with total mass $M_{\rm sim}=10^{28}$~g. The crust initially occupies a spherical shell with the outer radius $R=11$~km. It is initially static and stratified in hydrostatic equilibrium with a power-law index $q$: $\rho\propto x^q$ where $x=R-r$ is the depth below the stellar surface at a radius $r<R$. We have run simulations with $q=5$ and $q=3$. Our simulations are performed using a relativistic Lagrangian hydrodynamic code described in \cite{Lundman2018}, with some modifications. In particular, we use a non-uniform discretization in the mass coordinate $0<m<M_{\rm sim}$, which allows us to resolve well the dynamics of the low-density layers near the stellar surface. The simulated mass $M_{\rm sim}$ is discretized into $10^4$ subshells. 

\begin{figure}[t]
 % \begin{center}
\includegraphics[width=0.47\textwidth]{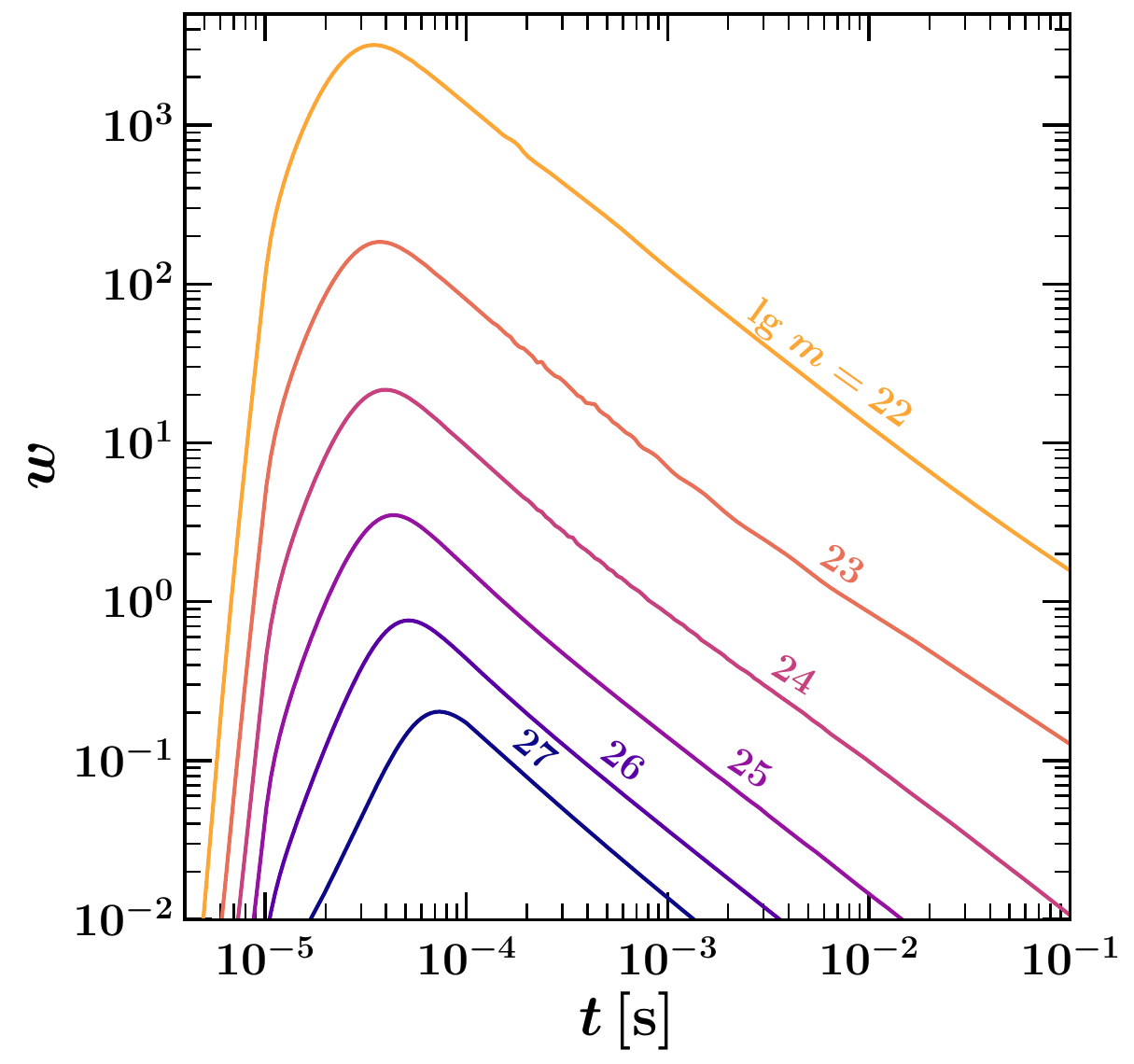} 
 % \end{center}
 \caption{Heating history of ablated surface layers. The dimensionless enthalpy $w=4P/\rho c^2$ is shown as a function of time $t$ for several layers with different Lagrangian mass coordinates $m$; $m$ is measured inward from the pre-ablation stellar surface. This sample ablation model assumes the initial hydrostatic stratification index $q=5$ and the geometric parameter of the neutrino-sphere $a=4$~km.}
 \label{w}
\end{figure}

\begin{figure}[t]
 % \begin{center}
\includegraphics[width=0.47\textwidth]{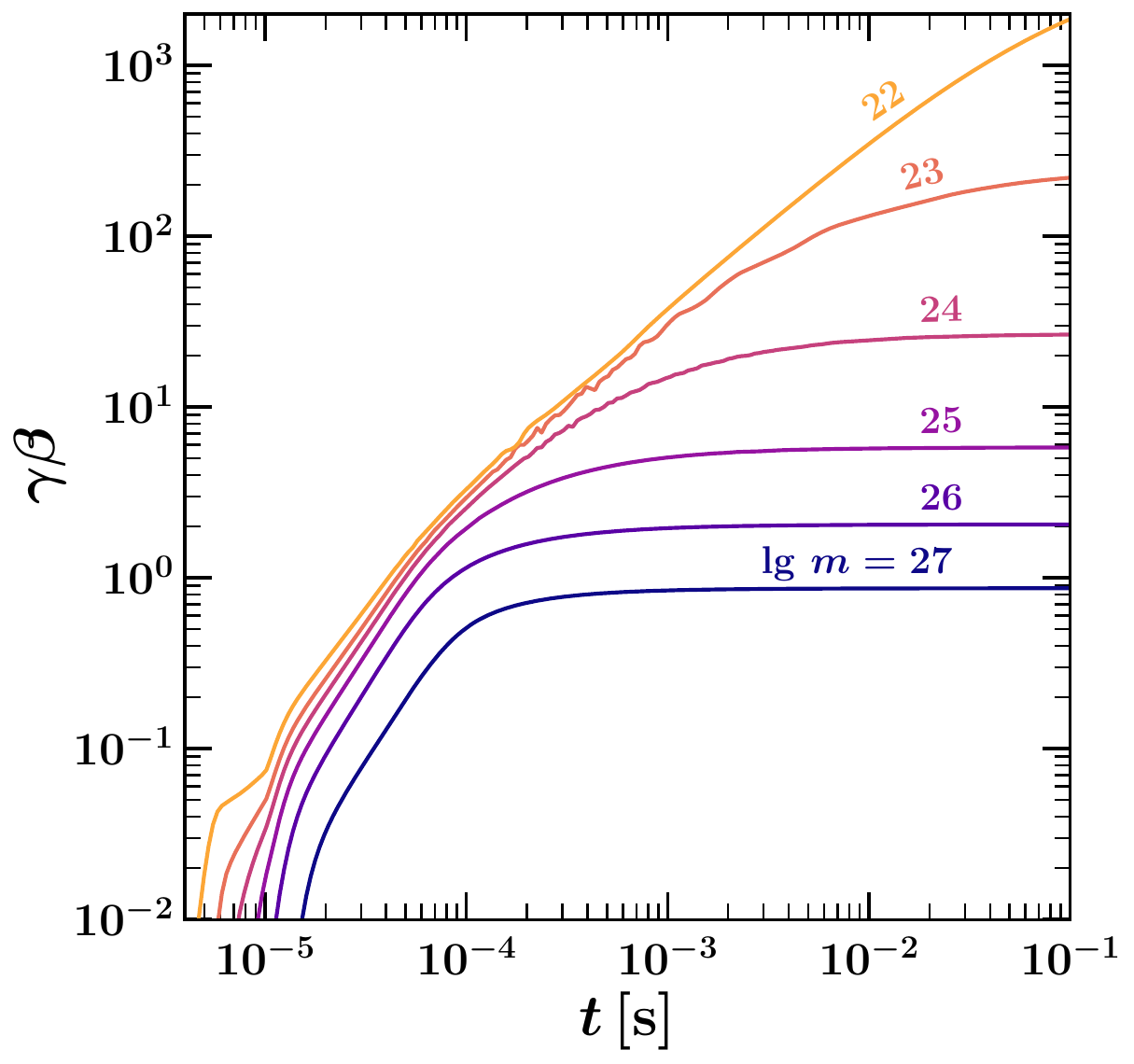} 
 % \end{center}
 \caption{Evolution of the dimensionless momentum $\beta\gamma$ of the same layers as in Figure~\ref{w}. 
 }
 \label{mom}
\end{figure}

\begin{figure}[t]
 % \begin{center}
\includegraphics[width=0.47\textwidth]{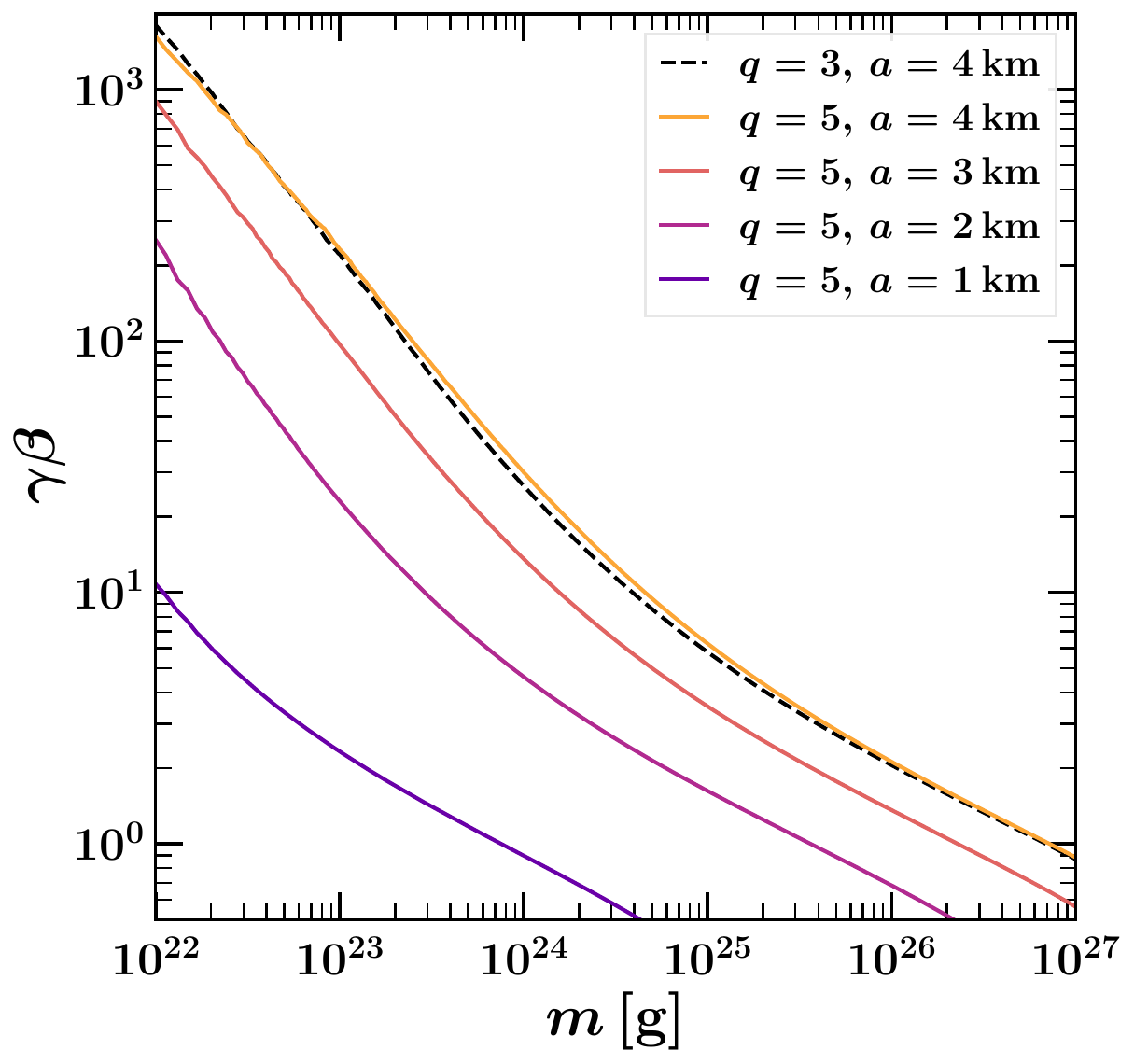} 
 % \end{center}
 \caption{
 Asymptotic 4-velocity $\gamma\beta$ of the relativistically ablated layers as a function of the Lagrangian coordinate mass $m$ measured inward (i.e. $m=0$ at the outer boundary of the crustal material). Our ablation simulation included only upper layers $m<10^{28}$~g, which are heated mainly via reaction $\nu\bar{\nu}\rightarrow e^+e^-$ (later ejecta emerge from deeper and denser layers heated mainly via neutrino absorption by nucleons). The results are shown for several models that have heating rates $\dot{Q}$ (see text) with different choices of the parameter $a$ that depends on the shape of the sandwich neutrino-sphere. The results are sensitive to $a$ and weakly depend on the initial stratification of the surface layers $\rho\propto x^q$ in the relevant range of $q$.}
 \label{mass}
\end{figure}

The results are shown in Figures~\ref{w}-\ref{mass}. One can see that the heat is deposited in the surface layers on the timescale $t\sim (3-7)\times 10^{-5}$~s. Then the layers leave the main heating zone, expand and cool, converting the accumulated heat to bulk kinetic energy. The relativistically ablated matter approaches its final (asymptotic) momentum $\gamma\beta\gg 1$ after a longer time, when it has lost its enthalpy through adiabatic cooling. 

Figure~\ref{mass} shows how much mass escapes with $\beta\gamma$ larger than a given value. The result is sensitive to the geometric parameter $a$. For instance, in the  most ``optimistic'' case of $a=4$~km, mass $m\approx 10^{26}$~g is ejected with $\gamma\beta>2$, and $m\approx 10^{25}$~g is ejected with $\gamma\beta>4$. One can also see that the distribution $\gamma\beta(m)$ extends to very high values of $\gamma\beta\sim 10^3$. This is expected, as at the onset of ablation the neutrino wave deposits comparable energy everywhere in the heating zone in the upper crust (or even above it), regardless of the local density of matter. As a result, the outermost layers of the neutrino-driven outflow form an ultra-relativistic fireball that freely expands with acceleration by the radiation pressure $\gamma\propto r$. The fireball Lorentz factor is limited to $\gamma\sim 10^3$, because at higher $\gamma$ the fireball becomes transparent to radiation and acceleration becomes inefficient.

%%%%%%%%%%% FIGURE %%%%%%%%%%%%%%%%%%
%%%%%%%%%%% FIGURE %%%%%%%%%%%%%%%%%%

%%%%%%%%%%% FIGURE %%%%%%%%%%%%%%%%%%
%%%%%%%%%%% FIGURE %%%%%%%%%%%%%%%%%%

%%%%%%%%%%% FIGURE %%%%%%%%%%%%%%%%%%
%%%%%%%%%%% FIGURE %%%%%%%%%%%%%%%%%%

%########################################################

\section{Early internal shock in the merger ejecta}
\label{bo}

This section describes the second mechanism for generating a self-similar ultra-relativistic envelope. It invokes a shock wave accelerating in the outer layers of the massive cloud around the merger remnant.

This mechanism is similar to shock breakout in supernovae, see e.g. detailed calculations in \cite{Tan2001} for spherical shocks, which are parallel to the surface of the supernova progenitor. These calculations were applied to merging neutron stars by \cite{Kyutoku2014}. 
We note that the merger shocks propagate obliquely to the neutron star surface. When the oblique shock approaches the upper crust, the sudden density drop will make the shock perpendicular to the surface, rather than parallel, and this effect introduces an upper cutoff on the velocity of the ejected material \citep{Matzner2013}. Therefore, it is unclear if a sufficient amount of ultra-relativistic ejecta can be produced by the merger shocks launched inside the neutron stars. Instead, below we focus on internal shocks in the large cloud around the merger remnant. Shock acceleration in the expanding cloud is more capable of ejecting a relativistic envelope.

Clouds around the merger remnants were predicted to have masses $\Mej\sim 10^{-2}M_\odot$ and expansion speeds $\vej/c\sim 0.1-0.3$ (e.g. \citealt{Bauswein2013,Siegel2017}). The cloud inferred from the kilonova in GW~170817 is somewhat more massive, $\Mej\sim 0.05M_\odot$.  An internal shock must develop in the cloud if matter ejected at later times has a higher speed. In particular, this is expected if the merger produces a magnetar --- a massive, differentially rotating neutron star which generates ultrastrong magnetic fields. Although the field dynamics in the merger is not fully understood, it is plausible that an ultrastrong field is generated and then rises to the stellar surface due to magnetic buoyancy \citep{Kluzniak1998}. Then a strong magnetosphere should form, and the neutrino-driven wind from the massive neutron star becomes centrifugally accelerated \citep{Mestel1987,Metzger2018}. This fast wind will drive a shock wave into the earlier, slower ejecta. \cite{Metzger2018} argued that such a wind could help explain the ``blue'' part of the kilonova emission in GW170817.

A shock with a velocity jump $\delta v$ crossing the cloud of mass $\Mej$ will dissipate energy 
\beq
\label{eq:E}
  \En\sim \frac{\Mej(\delta v)^2}{2}\approx 10^{51} \left(\frac{\Mej}{10^{-2}M_\odot}\right)
  \left(\frac{\delta v}{10^{10}{\rm  cm/s}}\right)^2 {\rm ~erg}.
\eeq
As long as the shock propagates deep inside the cloud, its speed remains mildly relativistic. When it reaches the outer layers of the cloud, where density is lower, the shock will accelerate and bring the outer layers to ultra-relativistic speeds. This process of shock breakout will inflate a self-similar relativistic envelope around the merger remnant.

Below we estimate the mass of the relativistic envelope and its structure, and illustrate it with a hydrodynamical simulation. Then we estimate the photon-to-baryon ratio in the inflated envelope.

\subsection{Previous results for shock breakout in static clouds}
\label{bo_prev}

Shock acceleration in non-relativistic hydrodynamics was studied in detail six decades ago (see \cite{Zeldovich1967} and references therein). Its relativistic version was proposed as a possible mechanism for outflows in GRBs \citep{Paczynski1998}.

\cite{Tan2001} provided analytical fits for relativistic mass ejection by shock breakout, which were tested against hydrodynamical simulations. In their simulations a shock emerges from an initially static star with density $\rho=0$ at the stellar surface, and a polytropic mass stratification with depth $x$ below the surface, $\rho\propto x^n$, with a typical $n=3$. The form of their analytical approximation is motivated by earlier results (obtained in the relativistic and non-relativistic limits), which are valid for more general density profiles. Therefore, similar ejecta are expected for shock breakout in clouds with different density distributions, as long as $\rho$ steeply drops in the outer layers of the cloud.

The main dimensionless parameter of the problem is the ratio $\En/\Mej c^2$.
For strong shocks in the merger ejecta, we expect
\beq
\label{eq:E}
   \tilde{\En}\equiv \frac{\En}{\Mej c^2}\sim 0.01-0.1,
\eeq
which corresponds to $\delta v\sim (0.1-0.5)c$ during the shock propagation inside the cloud, before the breakout.

The breakout problem has two parts: shock dynamics and subsequent expansion of the shock-heated fluid, with adiabatic cooling and bulk acceleration. The growth of the shock speed $\bs=\vs/c$ with decreasing density $\rho\ll \Mej/r^3$ is approximately described by
\beq
\label{eq:vsh}
    \gs\bs \approx A\,\tilde{\En}^{1/2} \left(\frac{\Mej}{\rho\, r^3}\right)^{\alpha},
\eeq
where $\gs=(1-\beta_s^2)^{-1/2}$ and $A\sim 1$ is a numerical factor. The power-law index $\alpha\approx 0.2$; its more accurate value is 0.187 when $\gs\bs\ll 1$ \citep{Zeldovich1967} and $\alpha=\sqrt{3}-3/2\approx 0.23$ when $\gs\bs\gg 1$ \citep{Johnson1971,Pan2006}. As $\bs$ grows with decreasing $\rho$, the dissipated energy per unit mass increases, however the energy density decreases. Most of the energy $\En$ is dissipated in the dense, heavy part of the ejecta, and only a fraction of $\En$ is delivered to the outer, low-density layers that eventually develop highly relativistic motion $\gamma\beta>1$. 

Figure~6 in \cite{Tan2001} shows how the ejecta energy is distributed over the asymptotic $\gamma\beta$ for several choices of $\tilde{\En}$. In particular, for $\tilde{\En}\sim 0.03$ they find that the ejecta with asymptotic $\gamma\beta>1$ carry the energy of $\En_1\approx 6\times 10^{-5}\Mej c^2$, and these ejecta have mass $m_1\sim 10^{-4}\Mej$. When a similar estimate is applied to the merger cloud, it gives $m_1\sim 10^{-6}M_\odot(\Mej/0.01M_\odot)$. Ejecta with yet higher $\gamma\beta\gg 1$ have a significantly smaller mass, e.g. $m\sim 10^{-6}\Mej$ for $\gamma\beta>3$. 

These estimates are sensitive to $\tilde{\En}$. One can see from Figure~6 in \cite{Tan2001} that a change of $\tilde{\En}$ from 0.03 by a factor of 3 (in either direction) changes $m_1$ by approximately two orders of magnitude. A crude estimate in the relevant parameter range may be written as
\beq
   m_1\equiv m(\beta\gamma>1)\sim 10^{-6}\,M_\odot \left(\frac{\tilde{\En}}{0.03}\right)^4 \left(\frac{\Mej}{0.01M_\odot}\right),
\eeq
\beq
   \frac{m(\beta\gamma>3)}{m(\beta\gamma>1)}\sim 3\times 10^{-3}.
\eeq
While \cite{Tan2001} considered a static star with a certain density profile, similar order-of-magnitude estimates apply to shocks in expanding clouds. In \Sect~4.2 we perform a detailed calculation for a sample cloud model and find the accurate distribution of the asymptotic four-velocity in the ejected envelope.

\subsection{Simulation of early shock breakout from an expanding cloud}
\label{bo_sim}

At the start of the simulation (time $t=t_0$), we specify the cloud parameters as follows. We place a spherical shell of mass $\Mej=10^{-2}M_\odot$ with the outer radius $R_c=2\times 10^9$~cm and the inner radius of $10^9$~cm. The outer half of the shell is expanding with a uniform speed $v_0=0.1c$, and the inner half is expanding with $v_0+\delta v$, where $\delta v/c=(2\tilde{\En})^{1/2}\approx 0.45$ corresponds to $\tilde{\En}=0.1$. 

We assume that the cloud was adiabatically cooled as it expanded from the merger remnant of radius $\sim 10^6$~cm, and we give the shell a low (insignificant) enthalpy $w=(U+P)/\rho c^2=10^{-3}$. The density profile of the shell is flat except near its outer edge, where density falls off exponentially on a scale $\dr=0.3 R_c$. We choose the moderate $\Delta r/R_c$ keeping in mind that ejecta from neutron star mergers are initially hot, and there is significant pressure in the cloud until it strongly expands and cools adiabatically. Even if the cloud was initially ejected with a sharp edge, the pressure drop in the outermost layers will accelerate them, creating a positive velocity gradient in the radial direction. It leads to stretching of $\dr$ and makes the density decline at the edge smooth and gradual. Therefore, freely expanding warm clouds in general cannot have  sharp edges. They are also generally expected to have a positive gradient of $v_0(r)$, so our assumption of $v_0(r)=const$ in the outer layers is a rather crude simplification of the expansion velocity profile.

%%%%%%%%%%% FIGURE %%%%%%%%%%%%%%%%%%
\begin{figure*}[t]
 \begin{center}
\includegraphics[width=0.9\textwidth]{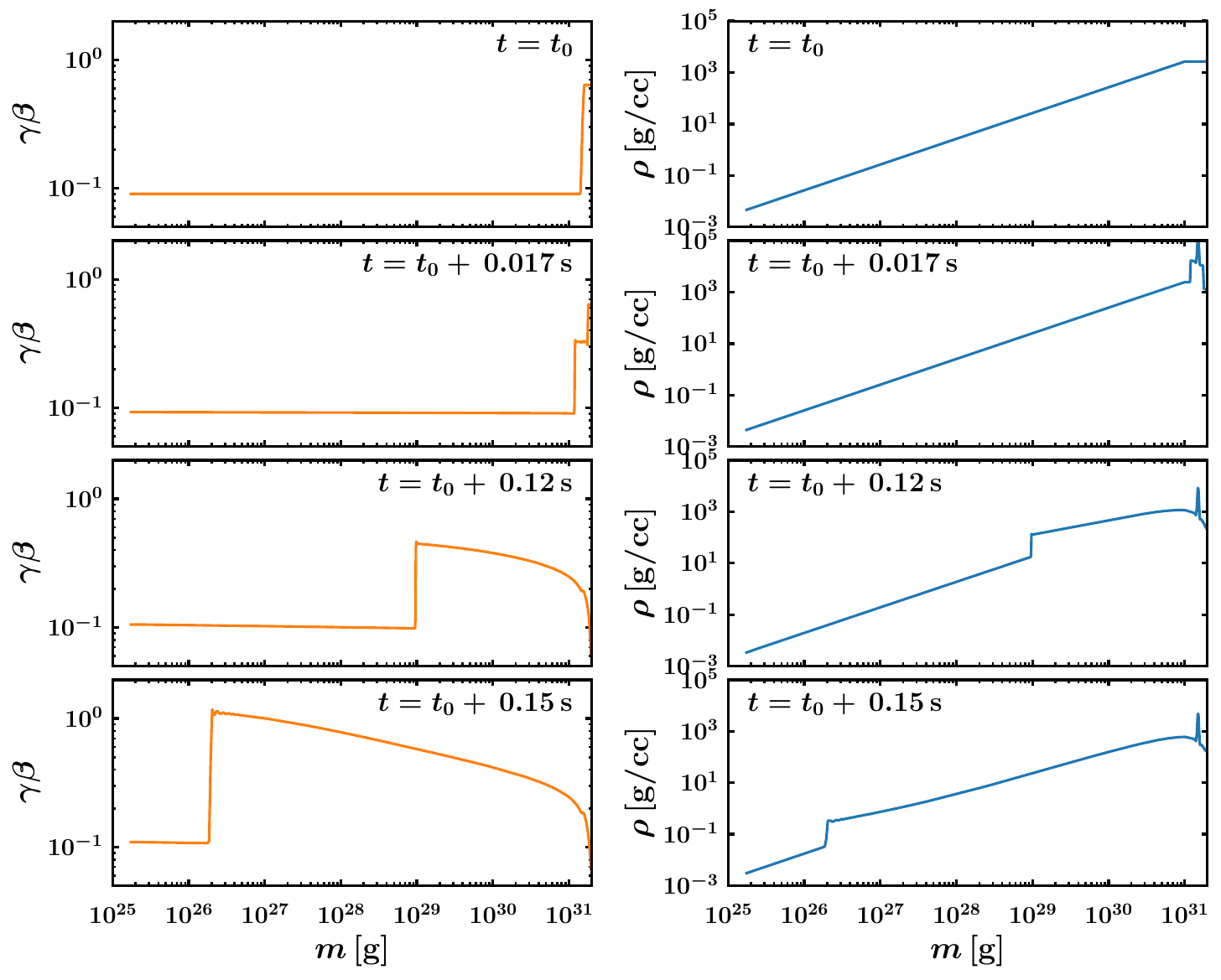}
 \end{center}
 \caption{Evolution of the fluid density profile $\rho(m)$ (right) and the 4-velocity profile $\beta\gamma(m)$ (left). The Lagrangian coordinate $m$ is measured inward; $m=0$ at the outer boundary of the simulated cloud and $m=2\times 10^{31\,}$g at the inner boundary.}
 \label{bo}
\end{figure*}
%%%%%%%%%%% FIGURE %%%%%%%%%%%%%%%%%%

Our initial condition with the velocity jump $\delta v$ inside the cloud of radius $\sim 10^9$~cm roughly corresponds to shock launching from the central object at time $\sim 1/2$~s after the merger. In the simulation, the jump immediately launches a forward shock in the middle of the cloud. There is also a reverse shock, however we are mainly interested in the dynamics of the forward shock, which propagates outward, reaches the low-density layers, and accelerates. Similar to the ablation simulation in \Sect~3, we use the Lagrangian mass coordinate $m$ counted from outside, and employ the Lagrangian relativistic hydrodynamics code of \cite{Lundman2018}. The non-uniform discretization in $m$ allows us to track the evolution of the entire massive shell $\Mej=0.01M_\odot$ while resolving the dynamics of the outer, low-$m$ layers, where the forward shock reaches highly relativistic speeds.

Figure~6  shows snapshots of fluid density $\rho(m)$ and four-velocity $\gamma\beta(m)$ in our simulation. It demonstrates the shock evolution and its effect on the structure of the cloud. As the forward shock enters the outer low-density layers and accelerates, it loses causal contact with the inner massive part of the cloud. The shock completely crosses the outer half of the cloud in $\sim 1/4$~s, and after this the shocked layers continue to expand with acceleration, converting heat to bulk kinetic energy. Then the ejecta become cold and ballistic.

\begin{figure}[t]
 % \begin{center}
\includegraphics[width=0.47\textwidth]{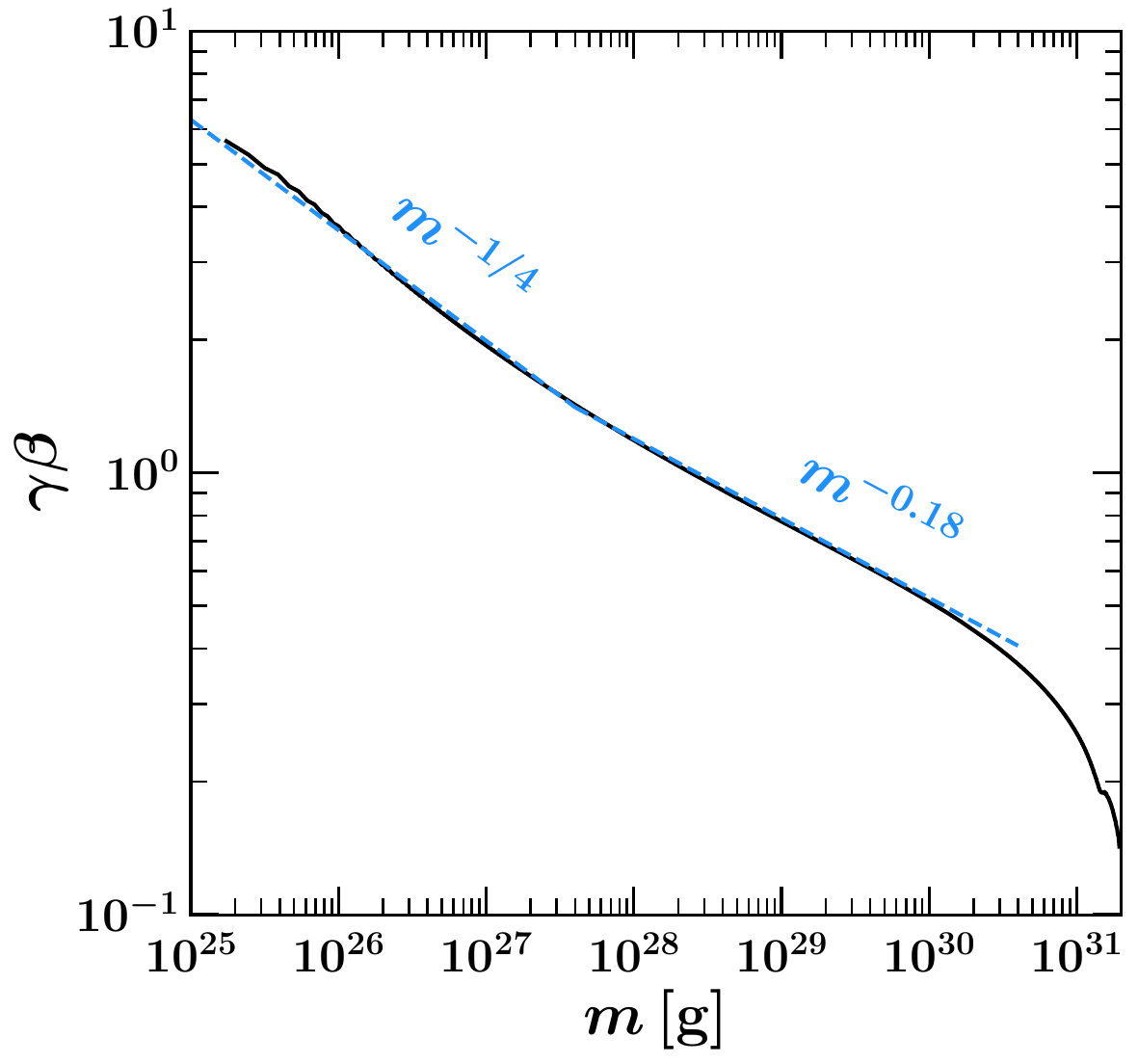}
 % \end{center}
 \caption{The final distribution of the four-velocity $\gamma\beta$ in the envelope launched by the internal shock breaking out from the cloud of massive ejecta. The Lagrangian coordinate $m$ is measured inward ($m=0$ at the outer boundary of the envelope). The cloud mass in the simulation was $\Mej=0.01M_\odot$, and its pre-shock expansion speed was $v_0=0.1 c$. The  shock was created by depositing an additional energy of $10^{51}$~erg in the inner half of the cloud (see text and Figure~6). The black curve shows the result of the numerical simulation, and the blue dashed line shows a fit by the broken power-law, with the break at $\gamma\beta=4/3$.}
 \label{bo_final}
\end{figure}

The final profile of $\gamma\beta$ as a function of the Lagrangian mass coordinate $m$ is shown in Figure~\ref{bo_final}. It may be approximately  described by power laws with different slopes $\psi$ in the regions $\gamma\beta\simlt 1$ and $\gamma\beta\gg 1$,
\beq
\label{eq:fit}
   \gamma\beta\approx \left(\frac{m}{m_1}\right)^{-\psi}.
\eeq
In our sample simulation with $\tilde{E}=0.1$, we find $m_1\approx 10^{-5}M_\odot$. The power-law slope is $\psi\approx 0.18$ at $m\simgt m_1$ and
$\psi\approx 1/4$ at $m\ll m_1$. 

In this sample simulation we assumed a flat pre-shock velocity profile $\beta_0=0.1$. More realistic expanding clouds have an increasing profile $v_0(r)$, shaped during the cloud formation near the central object. Changing the shape of $v_0(r)$ slightly changes the results, as long as $v_0\sim 0.1$. We also run models with a faster pre-shock speed $v_0=0.1-0.3$; then the profile of $v_0(r)$ significantly affects the final distribution of $\gamma\beta$ after shock breakout. Furthermore, the detailed shape of this distribution is affected by the initial density profile in the outer layers, and the initial enthalpy in the cloud. However, in all runs we found the final $\gamma\beta(m)$  qualitatively similar to that shown in Figure~7: a shallow power law at $\gamma\beta\simlt 1$, and a steeper power law at $\gamma\beta\gg 1$, with $\psi$ close to 1/4.

%%%%%%%%%%% FIGURE %%%%%%%%%%%%%%%%%%
%%%%%%%%%%% FIGURE %%%%%%%%%%%%%%%%%%

\subsection{Photon-to-baryon ratio in the envelope}
\label{bo_rad}

The early shock breakout described above occurs at a modest distance $\rb$ from the merger. This distance depends on the timescale of launching the shock, $t_{\rm sh}$, which may be related to the magnetar formation in the center; we assume $t_{\rm sh}<1$~s. During this time the ejecta expands to a radius $r\sim v_0 t_{\rm sh}<10^{10}$~cm, and the shock breaks out at a similar radius. It emits a burst of radiation when it reaches the ejecta photosphere, however this burst is too weak to be observed, as discussed below. Instead, the main radiative effect of the early internal shock is an increase of the photon number trapped in the ejecta. The large photon number will play a role later, when a new explosion from the merger remnant energizes the envelope at large radii $r\simgt 10^{12}$~cm, and a detectable GRB is emitted (\Sect~\ref{GRB}).

First let us estimate photon number carried by the massive cloud without an internal shock. The ejection of mass $\Mej$ in time $\tej$ implies a characteristic mass outflow rate $\dot{M}\sim\Mej/\tej=2\times 10^{31}(\Mej/0.01M_\odot)(\tej/1{\rm ~s})^{-1}$. The ejection radius $R_0\sim 20$~km (the size of the merger remnant) and expansion speed $\beta=v/c\sim 0.1-0.3$ determine a characteristic density $\rho_0\sim \dot{M}/4\pi R_0^2\, v\sim 10^8\dot{M}_{31}/\beta_{-1}$~g~cm$^{-3}$. Note that the expansion speed requires an initial enthalpy per unit rest mass $w_0\sim \beta^2/2$ and hence a thermal energy density at the base of the outflow $U_0=(3/4)w_0\rho_0 c^2\sim 10^{27}\dot{M}_{27}\beta_{-1}$~erg~cm$^{-3}$. This rough, order-of-magnitude estimate is sufficient to evaluate a characteristic temperature from $aT_0^4\sim U_0$: $kT_0\sim 1\,\dot{M}_{31}^{1/4}\beta_{-1}^{1/4}$~MeV. A more accurate estimate takes into account that at temperatures $\sim 1$~MeV $e^\pm$ pairs make a contribution to $U_0$ comparable that of photons; then $T_0$ is reduced by the factor of $(11/4)^{-1/4}\approx 0.8$.

The photon number density is $n_\gamma=U_0/2.7kT_0$, and the baryon number density  is $n_b=\rho_0/m_p$. The photon-to-baryon ratio in the massive cloud is then given by 
\beq
\label{eq:number}
   \frac{n_\gamma}{n_b}\approx \frac{(3/4)w_0m_pc^2}{2.7kT_0} \sim 10\, v_{10}^{7/4} \dot{M}_{31}^{-1/4}
      \quad {\rm (without~shock)}.
\eeq
The internal energy and entropy of the cloud are dominated by radiation, and during its adiabatic expansion the photon number remains approximately constant, because it is proportional to entropy. The conservation of the photon and baryon numbers implies that their ratio remains unchanged as the cloud expands.

The simple model of adiabatic expansion can be refined by including two effects. First, photon number is increased by a moderate factor $\sim 2$ when the temperature decreases below $m_ec^2=511$~keV and the $e^\pm$ pairs annihilate into photons. Second, when the cloud temperature drops to $kT_{\rm rec}\sim 150-200$~keV, free nucleons recombine into $\alpha$-particles, releasing 28~MeV per $\alpha$-particle (see \cite{Beloborodov2003} for a discussion of $T_{\rm rec}$ in GRBs). The number of photons generated by the recombination may be estimated as 
\beq
\label{eq:rec}
  \frac{n_\gamma}{n_b}\sim 2Y_e\,\frac{7\,{\rm MeV}}{2.7kT_{\rm rec}}\sim 10,
\eeq
where $Y_e<0.5$ is the proton-to-baryon ratio in the cloud.
The produced photon number is small, comparable to that in \Eq~(\ref{eq:number}) and orders of magnitude smaller than $n_\gamma/n_b$ generated by the shock discussed below. 
  
 Deep inside the massive cloud the photon number may be calculated taking into account the synthesis of heavy neutron-rich elements and their $\beta$-decay at a later stage. However, nucleosynthesis of elements heavier than helium is inefficient in the outermost, low density layers of main interest for us, and so we do not include this effect here.

Let us now consider the production of photon number by an internal shock  in the cloud. The shock boosts $n_\gamma/n_b$, because it generates entropy and radiation at radii $r\sim v_0t_{\rm sh}\gg R_0$. The jump conditions for a shock with speed $v_s=\beta_s c$ give the generated energy density $U\approx 2\gs^2\bs^2\rho c^2$, where $\rho$ is the fluid density ahead of the shock. We are interested in the outer layers of the cloud, $m\ll \Mej$, where the shock accelerates to $\gs\bs\simgt 1$. The cloud density $\rho(m)$ may be estimated as follows. Each layer of the cloud at radii of interest expands ballistically, with some positive velocity gradient, and the expansion is homologous, with $\rho(m,t)\propto t^{-3}\propto r^{-3}$. The parameter $\xi=r/\dr$ (which describes the sharpness of the density decline at the outer edge) freezes for homologous ballistic expansion, and so the density of the pre-shock layers may be written as
\beq
\label{eq:rhom1}
  \rho(m,t)\approx \frac{\xi\, m}{4\pi r^3(t)}, \qquad \xi=\frac{r}{\dr}\simgt 1.
\eeq 
Using this equation for $\rho$ and \Eq~(\ref{eq:vsh}) for $\gs\bs$, one can express $U\approx 2\gs^2\bs^2\rho c^2$ as a function of the mass coordinate of the propagating shock, $m$. This gives the following estimate for the post-shock energy density,
\beq
   U(m)\sim 2\gs^2\vs^2(m)\,\rho(m) \sim  \frac{\tilde{\En}c^2}{r^3}\left(\frac{\xi\mmm}{4\pi}\right)^{0.6} \Mej^{0.4}.
\eeq
Here $r\sim \rb$ is the shock breakout radius. Assuming that the energy density $U$ is quickly thermalized, we can estimate the blackbody radiation density $aT^4\approx U$, which gives the temperature
 \beq
 \label{eq:Tsh}
    kT\sim 40 \,m_{26}^{0.15} \,r_{9}^{-3/4} \left(\frac{\Mej}{0.01M_{\odot}}\right)^{0.1}
       \left(\frac{\tilde{\En}}{0.1}\right)^{1/4}       {\rm keV}.  
\eeq
The generated photon number per nucleon in the region where the shock accelerates to $\gs\gg 1$ is given by
\beq
\label{eq:ratio}
   \frac{n_\gamma}{n_b} \sim \frac{\gs\, m_pc^2}{\sqrt{2}\, 2.7kT}
     \sim 10^4 m_{26}^{-0.35} r_{9}^{3/4}  
     \quad {\rm (after~shock).}
\eeq
In this last estimate we have omitted the weak dependence on $\Mej$ and $\tilde{\En}$.

The assumption of quick thermalization of radiation in a heated flow can be verified as follows
 (e.g. \citealt{Levinson2012,Beloborodov2013}). Radiation must relax to a Planckian spectrum with the photon density $n_\gamma=U/2.7kT$ if the plasma efficiently emits photons. The two main processes of photon production by the thermal plasma is double Compton scattering and bremsstrahlung. In particular, double Compton scattering occurs with rate $\dot{n}_{\rm DC}\approx 0.1\,n_\gamma n\,\sT c\,\Theta^{2}$, where $n$ is the electron/positron number density and $\Theta=kT/m_ec^2$. The number of photons produced during the expansion timescale for the post-shock plasma is $n_{\rm DC}\approx \dot{n}_{\rm DC}\, r/c$. It becomes exponentially large, ensuring thermalization, if  
\beq
\label{eq:DC}
  \frac{n_{\rm DC}}{n_\gamma}\approx 0.1 n\sT r\,\Theta^2=0.1\,\xi\,\tauT\,\Theta^2 \gg 1.
\eeq 
Here $\tauT\approx n\sT\dr$ is the scattering optical depth of the ejecta outside the current shock radius. It is related to the mass coordinate $m$ by
\beq
\label{eq:tau}
   \tauT=\frac{\sT Y_e m}{4\pi r^2 m_p}\approx 3\times 10^4\, Y_e\, m_{26}\, r_{10}^{-2},
\eeq
where $\sT$ is Thomson cross section, and $Y_e=n_p/n_b$ is the proton-to-baryon ratio. For simplicity, we neglected the possible presence of $e^\pm$ pairs (which could only increase the photon production rate). One can see that the condition~(\ref{eq:DC}) is satisfied, and so the shock-generated radiation is thermalized in the shells of interest $m>10^{25}$~g, as long as the shock crosses the cloud before it expands to $r\sim 10^{10}$~cm.

\subsection{Free neutrons}

An important feature of the merger ejecta is their neutron-rich composition $Y_e<0.5$. The baryons at the base of the outflow are initially free nucleons, predominantly neutrons. As the matter expands and cools, the nucleons recombine into $\alpha$-particles, and the neutron excess implies some leftover free neutrons. Deep inside the massive cloud  most of the free neutrons become locked into heavy, neutron-rich nuclei after $\sim 1$~s of expansion (e.g. \citealt{Metzger2010}). However, this process is less efficient in the outer layers of main interest here, $m\sim 10^{25}-10^{27}$~g, because their density is well below the typical density inside the massive cloud.

The free neutrons and ions are still well coupled by frequent nuclear collisions, so to a first approximation one can treat them as a single fluid. However, this approximation is not valid on small scales comparable to the shock thickness, and the drift of neutrons relative to the ions changes the shock dissipation mechanism \citep{Beloborodov2017}. In the absence of free neutrons, the shock is mediated by radiation and has a thickness comparable to the photon free path. In the presence of free neutrons, the shock is partially mediated by neutrons, which have much longer free paths. In addition, the neutron-ion collisions around the shock cause spallation of $\alpha$-particles \citep{Belyanin2001,Beloborodov2003}.

When the shock becomes highly relativistic, $\gs\bs\simgt 1$, the neutron-ion collisions in the shock become inelastic and generate pions. The pions immediately decay into ultra-relativistic leptons and generate a nonthermal inverse Compton cascade. In the presence of a magnetic field, the cascade would be capable of producing a significant photon number through synchrotron emission \citep{Beloborodov2010,Vurm2016}. However, the envelope magnetization is likely low --- both mechanisms of the envelope ejection (\Sects~3 and 4) suggest that it expands outside the magnetic fields generated by the merger. This suggests weak synchrotron emission by the cascade from neutron collisions.

\subsection{Photospheric emergence of the early shock}

We also note that the early shock breakout radiates little energy when it reaches the photosphere of the cloud. It does not produce a detectable burst, because of the modest emission radius $r\sim\rb<10^{10}$~cm. The photosphere is located in the outermost layers with mass 
\beq
\label{eq:mph}
    \mph\sim \frac{4\pi r^2 m_p}{Y_e \sT}
    \approx 3\times 10^{21}\, \frac{r_{10}^2}{Y_e} {\rm ~g}, 
\eeq
which we estimated from \Eq~(\ref{eq:tau}) by setting $\tauT\sim 1$.\footnote{Density of the photospheric layers is low compared with the inner parts of the cloud. Therefore heavy nuclei are not synthesized in these layers, and there is no bound-free absorption of photons, so we assume Thomson opacity.} Since $\mph$ is so small, the internal shock must accelerate to a high Lorentz factor $\gs(\mph)\sim 10$ as it reaches the photosphere. The energy of the shocked photospheric layers is given by
\beq
  \En_\star\sim \gs^2(\mph)\, \mph c^2.
\eeq
It is 2-3 orders of magnitude smaller than the energy of GRB~170817A, and so the early shock breakout is hardly capable of emitting detectable radiation. The large number of photons produced by the shock inside the cloud remain trapped by the huge optical depth and experience strong adiabatic cooling.

\medskip

%#################################################################

\section{GRB production in the envelope}
\label{GRB}

The main conclusion from the preceding sections is that the merger GW~170817 likely ejected  a low-mass, opaque envelope expanding with a stratified Lorentz factor $\gamma(m)\gg 1$. The cold ballistic envelope becomes capable of emitting a GRB only if it is reheated by some dissipation process. A simple way to accomplish this is to drive a new shock wave. Therefore, below we consider a scenario where the merger remnant produces a delayed explosion (e.g. \citealt{Gottlieb2018}). In particular, if the remnant is a super-massive neutron star with a limited lifetime (\citealt{Lipunova1998}), the explosion may be associated with its collapse. The collapse is promoted by the generation of ultra-strong magnetic fields and loss of differential  rotation, as well as by cooling due to neutrino emission, on a timescale of a few seconds. Then the nascent spinning black hole launches powerful, ultra-relativistic, magnetized jets.

Compared with the pre-collapse massive neutron star (the magnetar), the black hole is much more capable of launching the jets. During the collapse, the source of the baryonic wind polluting the magnetosphere of the magnetar disappears behind the event horizon. At the same time, the accretion disk of the merger debris continues to sustain a strong magnetic field threading the black hole. The Poynting flux from the black hole (of radius $\sim 5$~km) may exceed that from the magnetar, because it is more compact than the magnetar and is spinning faster. These conditions are favorable for formation of an ultra-relativistic jet, which is collimated by the surrounding slower ejecta.

\begin{figure*}[t]
 \begin{center}
\includegraphics[width=0.8\textwidth]{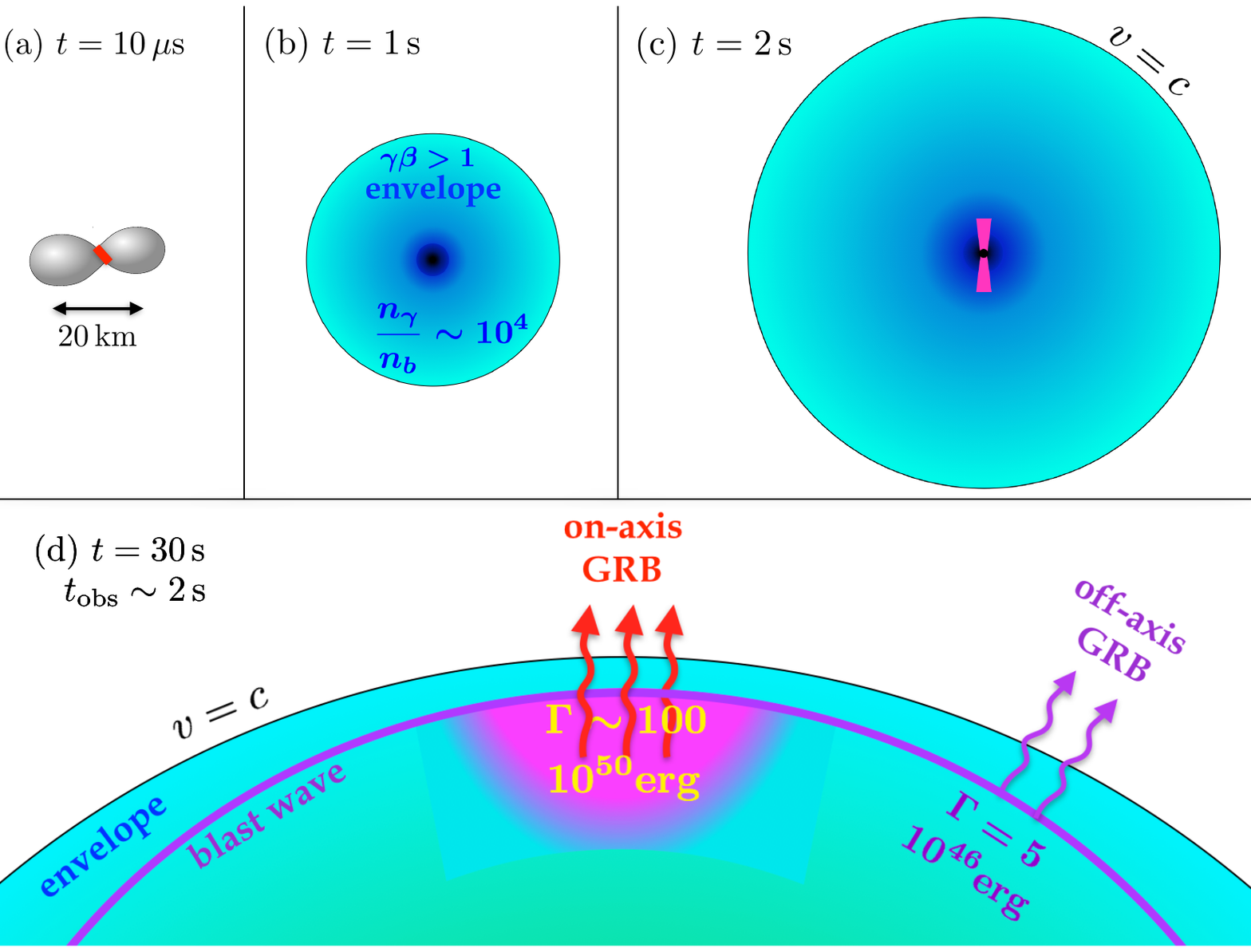}
 \end{center}
 \caption{Schematic summary of the explosion model.
(a) Two neutron stars begin to merge ($t\sim 10^{-5}$~s). At this stage, a hot sandwich (red) forms between the stars, and its powerful neutrino emission ablates the neutron star crust.
(b) The merger remnant (a magnetar) is engulfed by its ejecta, shown at time $t=1$~s. The ejecta are pictured schematically as a spherical cloud; real ejecta may not be spherical. The dense cloud of mass $M_{\rm ej}\simgt 10^{-2}M_\odot$ (dark blue) is expanding with speed $v\sim (0.1-0.3)c$. It will emit the kilonova a day later. The dense cloud is surrounded by the large ultra-relativistic envelope ($\gamma\beta>1$) of mass $m\sim 10^{-7}M_\odot$ (cyan). It was ejected by an internal shock wave that broke out of the cloud at $t<1$~s. The shock also loaded the envelope with a large number of photons $n_\gamma/n_b\sim 10^4$, which are trapped and adiabatically cooled. The envelope will continue to expand ballistically and homologously. The profile of its Lorentz factor $\gamma(m)$ diverges toward the outer edge, and the radial scale of its density variation is compressed as $\Delta r\sim r/\gamma^2$. 
(c) The magnetar has collapsed into a black hole at $t\simlt 2$~s, and a pair of powerful jets (bright magenta) have been launched by the accreting black hole. The jets will chase the outermost, ultra-relativistic layers of the envelope for a long time.
(d) The blast wave from the jet has expanded far into the ultra-relativistic envelope at all polar angles $\theta$. The radius of its photospheric emergence depends on $\theta$, because its energy and Lorentz factor $\Gamma$ decrease with $\theta$. 
By $t\sim 30$~s ($\tobs\sim t/2\gamma^2\sim 2$~s) the blast wave has reached the photosphere at $\theta\sim 0.5$, the viewing angle of GW~170817. The indicated characteristic energies of $10^{50}$~erg and $10^{46}$~erg are the outflow energies at $\theta\sim 0.1$ (the jet core) and $\theta\sim 0.5$. The corresponding isotropic equivalent of the energy is larger by the factor of $\sim (1-\cos\theta)^{-1}$, which gives $\sim 10^{52}$~erg for the jet core, and $\sim 10^{47}$~erg at $\theta\sim 0.5$. An ultra-bright on-axis burst is emitted by the jet plasma (bright magenta), which is more energetic than the blast wave ahead of it. The on-axis burst is powered by variable dissipation inside the jet and may have a complicated light curve. The off-axis burst is emitted by the blast wave in the envelope; its $\gamma$-ray light curve has a single peak and is followed by a soft X-ray tail.
 }
 \label{schematic}
\end{figure*}

Our proposed model for GRB production is schematically summarized in Figure~\ref{schematic}. The jets first propagate inside the massive cloud and then expand into the large ultra-relativistic envelope. The forward shock from the jet (and its cocoon in the cloud) forms a blast wave which initially expands forward and sideways around the jet. Simulations by \cite{Duffell2018} suggest that the blast wave will be launched into the outer envelope if the jet itself is successful, i.e. if it exits the massive cloud. At later times the blast wave shape becomes nearly spherical as it travels with almost speed of light and has the radius $r\approx ct$. The jet must be strongly collimated, as required by the late afterglow observations of GRB~170817A \citep{Lazzati2018,Granot2018,Lamb2018,Mooley2018b}. Therefore, the blast wave in the envelope has an anisotropic power, however it is less beamed than the jet.

When viewed at large angles from the polar axis, the explosion emission will be dominated by the blast wave in the envelope rather than the jet itself. By contrast, when viewed on-axis, the jet kinetic energy will strongly dominate over the energy of its forward shock in the envelope, and the observer will see a much brighter beamed GRB emitted by the jet plasma with $\Gamma\simgt 10^2$.

Below we focus on the off-axis GRB expected from the blast wave in the envelope, and compare it with GRB~170817A. We will assume that the jet launches the blast wave into the envelope at a time comparable to one second after the merger.

%%%%%%%%%%% FIGURE %%%%%%%%%%%%%%%%%%
%%%%%%%%%%% FIGURE %%%%%%%%%%%%%%%%%%

\subsection{Homologous density profile}

Let us first evaluate the radial structure of the relativistic envelope before the blast wave from the jet. The envelope structure can be calculated in the ballistic approximation. The picture becomes particularly simple at long times/large radii: the envelope may be thought of as a sequence of shells ejected at $r\approx 0$ and $t\approx 0$ with different speeds, so that a shell with mass coordinate $m$ and velocity $v(m)$ has the radius,
\beq
\label{eq:r}
   r(m,t)\approx v(m)\,t.
\eeq
A shell $dm$ has the thickness $dr=t\,|dv/dm|\,dm$, and hence the envelope has the following density distribution (measured in the fixed lab frame), 
\beq
\label{eq:rholab}
   \rho_{\rm lab}(m,t)\approx \frac{dm}{4\pi r^2 dr}\approx \frac{1}{4\pi v^2\,t^3\,|dv/dm|}.
\eeq
The proper density $\rho=\rho_{\rm lab}/\gamma$ is then given by
\beq
\label{eq:rho}
  \rho(m,t)\approx \frac{\gamma^2}{4\pi \beta^2 c^3t^3} \left|\frac{d(\gamma\beta)}{dm}\right|^{-1},
\eeq
where we have used the relation $d(\gamma\beta)=\gamma^3d\beta$. 

In particular, for the four-velocity distribution of the form $\gamma\beta=(m/m_1)^{-\psi}$ we find
\beq
   \rho(m,t)\approx \frac{m_1}{4\pi \beta^4c^3t^3\psi}\left(\frac{m}{m_1}\right)^{1-\psi}.
\eeq
The mass stratification of the ultra-relativistic homologous envelope is such that the radial thickness of an outer shell $m$ occupies a radial thickness of $\Delta r/r\sim \gamma^{-2}(m)$. The outer (smaller) $m$ occupies a progressively smaller $\Delta r\sim m\,(dm/dr)^{-1}$ because it moves faster and has a stronger relativistic compression factor $\gamma^{-2}$.

\subsection{Blast wave emergence at the photosphere}

The edge of the envelope has a diverging Lorentz factor $\gamma$, and so it is out of reach for the blast wave. Effectively, the blast wave propagates in an unbound medium, resembling the shocks in external media that produce GRB afterglows. Unlike the standard afterglow model, here the medium is opaque, and the shock is mediated by radiation. Furthermore, the blast wave is {\it accelerating}, because the envelope density is steeply decreasing with radius and time, and its Lorentz factor $\gamma$ increases with radius. The growth of the blast wave Lorentz factor $\Gamma$ gradually allows it to catch up with faster (and less massive) outer shells of the envelope. Eventually, the blast wave emerges at the photosphere of the envelope and produces a pulse of observed emission.

The photosphere of the homologously expanding envelope is located where the column number density of protons is $\sim \sT^{-1}$.  The photosphere mass coordinate $\mph$ and radius $\Rph$ are related by
\beq
\label{eq:mph1}
  \mph=\frac{4\pi \Rph^2 m_p}{Y_e\sT}.
\eeq
Radiation produced by the blast wave begins to escape to a distant observer when the shock mass coordinate $\ms(t)$ approaches $\mph(t)$. 

Relativistic shocks in a photon-rich medium are capable of creating copious $e^\pm$ pairs (\citealt{Beloborodov2017,Lundman2018,Ito2018}). This effect tends to prolong the photospheric emergence of the shock and delay its transition to complete transparency.
However, the shock in the envelope described above will have a mildly relativistic jump, and pair creation will not be so efficient (we leave this for future study).  In any case, the optical depth of pairs created by the shock is limited to $\sim 1$ by $e^\pm$ annihilation \citep{Beloborodov2017}. Therefore,  radiation will begin to escape the blast wave when $\ms=\mph$ even if pair creation is strong.

The observed delay of the shock appearance at the photosphere $\Rph$ is given by 
\beq
\label{eq:tph}
  \tph=\tobs(\Rph)\approx \frac{\Rph}{2c\gph^2}, \quad 
  \gph\equiv\gamma(\mph)\approx \left(\frac{\mph}{m_1}\right)^{-\psi}.
\eeq
Hereafter we assume that the relativistic envelope was inflated by the mechanism described in \Sect~4, and use \Eq~(\ref{eq:fit}).
Combining \Eqs~(\ref{eq:mph1}) and (\ref{eq:tph}) we find 
\beq
   \gph^{(1+4\psi)/\psi}=\frac{\sT Y_e m_1}{16\pi m_pc^2 \tph^2}.
\eeq
Note that $\psi/(1+4\psi)=(1/8)\pm 0.01$ in the relevant range of $0.2<\psi<0.3$, and so
\beq
\label{eq:gph}
   \gph\approx \left(\frac{\sT Y_e m_1}{16\pi m_pc^2 \tph^2}\right)^{1/8}
     \approx 3\,\tph^{-1/4}\,m_{1,27}^{1/8},
\eeq
where the observed time of the photospheric shock breakout $\tph$ is expressed in seconds. Then we also find $\Rph$ from \Eq~(\ref{eq:tph}),
\beq
\label{eq:Rph}
   \Rph\approx 5\times 10^{11}\,\tph^{1/2}\, m_{1,27}^{1/4}  {\rm~cm},
\eeq 
and the mass of the photospheric layers $\mph$ (\Eq~\ref{eq:mph1}),
\beq
  \mph \approx 2\times 10^{25}\,\tph\, m_{1,27}^{1/2} {\rm ~g}.
\eeq

The shock power depends on the jump of the fluid Lorentz factor, 
\beq
  \ggr\equiv \frac{\Gamma}{\gamma}.
\eeq
The shock jump conditions give the relative velocity between the downstream ($\Gamma$) and the upstream ($\gamma)$: $\brel=(\ggr^2-1)/(\chi^2+1)$, and the corresponding Lorentz factor is $\grel=(1-\brel^2)^{-1/2}=(\ggr^2+1)/2\ggr$. The energy dissipated by the shock as it crosses the photospheric layers of mass $\sim\mph$ is given by 
\begin{eqnarray}
\label{eq:Eph}
\nonumber
   \Eph &\approx& \frac{(\ggr_\star-1)^2}{2}\,\gph\,\mph c^2 \\
           & \approx& 3\times 10^{46}\,(\ggr_\star-1)^2\, \tph^{1/2}\,m_{1,27}^{5/8} {\rm~ erg}.
\end{eqnarray}

The photospheric shock breakout radiates a pulse of radiation with energy $\sim\Eph$ and the characteristic peak duration $\Delta\tobs\approx\Delta\tph\approx \Rph/2\Gph^2c$ determined by the Lorentz factor $\Gph=\Gamma(\mph)$ of the post-shock plasma. The peak width can be compared with its arrival time $\tobs\approx\tph$, 
\beq
\label{eq:width}
   \frac{\Delta\tobs}{\tobs}= \frac{\Delta\tph}{\tph} \approx \frac{\gph^2}{\Gph^2}
   \equiv\ggr_\star^{-2}.
\eeq
Thus, the relative width of the observed peak, $\Delta\tph/\tph$, can be used as a proxy for the shock strength at the photosphere.

Elsewhere we describe in more detail the blast wave propagation through the homologous relativistic envelope, and evaluate emission expected at $\tobs\gg\tph$, after the main peak of the off-axis GRB. It is produced by the deeper shells $\ms\gg \mph$, which were heated by the blast wave at smaller radii $\rs\ll\Rph$ and optical depths $\tau_s\gg 1$. The heated opaque shells behind $\mph$ release their radiation with a delay $\Delta\tobs(\ms)$, when they expand to transparency. This delayed emission is partially thermalized and adiabatically cooled, and so it is much softer than the GRB peak.

\subsection{Comparison with GRB~170817A}

The above predictions can be compared with observations of GRB~170817A. Its arrival time was $\tph\approx 1.7$~s and the main pulse had a width $\sim 3$ shorter than $\tph$, which implies $\ggr_\star\sim \sqrt{3}$ according to \Eq~(\ref{eq:width}). Substituting these values to \Eqs~(\ref{eq:Eph}) we find
\beq
  \Eph\sim 2\times 10^{46}\, m_{1,27}^{5/8} {\rm~erg}. 
\eeq
One can see that the envelope with $m_1\sim 3\times 10^{27}$~g (which is in the expected range for the envelope model in \Sect~4) is consistent with the observed energy of the main peak of GRB~170817A, $\En\sim 4\times 10^{46}$~erg \citep{Goldstein2017}. The corresponding mass of the photospheric layers is $\mph\sim 6\times 10^{25}$~g.

Furthermore, from \Eqs~(\ref{eq:gph}) and (\ref{eq:Rph}) we find that the pre-shock ejecta at the photosphere had Lorentz factor 
\beq
  \gph\approx 3.
\eeq
and the blast wave broke out at radius 
\beq
  \Rph\approx 10^{12} {\rm~cm}. 
\eeq 
The Lorentz factor of the radiating plasma immediately behind the shock is 
\beq
  \Gph=\ggr_\star\gph\approx 5.
\eeq 

The observed spectrum of the initial pulse peaked at $\Epk\sim 10^2$~keV, which roughly corresponds to the average photon energy $\bar{E}\sim 10^2$~keV (the detailed shape of the spectrum of GRB~170817A is uncertain, because of poor photon statistics).  This should be compared with the average energy of photons emitted by the blast wave at $\Rph$,
\begin{eqnarray}
\nonumber
  \bar{E}_\star &\approx& \frac{\Gs\grel m_pc^2}{n_\gamma/n_b}=\frac{(\ggr_\star-1)\,\gph m_pc^2}{2n_\gamma/n_b} \\
  &\approx & 100\,\left(\frac{n_\gamma/n_b}{10^4}\right)^{-1}{\rm ~keV}.
\end{eqnarray}
Thus, the observed $\Epk$ is consistent with the photon-to-baryon ratio $n_\gamma/n_b\sim 10^4$ expected in the envelope described in \Sect~4, see \Eq~(\ref{eq:ratio}). 

GRB~170817A was also reported to have a soft tail of emission after the main peak. In an accompanying paper we study soft emission expected after the blast wave breaks out of the opaque relativistic envelope. It has a decreasing luminosity $\Lobs(\tobs)$ and a decreasing average photon energy $\bar{E}(\tobs)$. However, quantitative tests of the tail prediction are difficult for GRB~170817A, because its tail  is barely detected and its properties are poorly known and still debated (cf. \cite{Burgess17}).

%##############################################

\section{Discussion}

The observed timing of GRB~170817A and its luminosity implies ultra-relativistic expansion of the gamma-ray source, $\Gamma\simgt 5$ (Figure~1). This constraint shows that neutron star mergers eject ultra-relativistic outflows at large angles from the rotation axis, at least up to $\theta\sim 20-30^\circ$ (the viewing angle for GW~170817). 

We have argued that this broad ultra-relativistic outflow has the form of a self-similar envelope expanding from the center with a stratified Lorentz factor. This picture follows from our investigation of a possible mechanism for ultra-relativistic ejecta. Both mechanisms described in \Sects~3 and 4 inflate a self-similar envelope with a Lorentz factor profile increasing outward. This envelope contains significant mass and is opaque. In particular, the ultra-relativistic envelope ejected by a magnetar shock (Section~4) can have mass exceeding $10^{-7}M_\odot$. It is sufficient for inflating the GRB photosphere to radii $r\sim 10^{12}$~cm.

A plausible scenario for producing gamma-rays invokes a delayed explosion from the merger remnant. The explosion launches a blast wave into the inflated envelope, which eventually emerges at its photosphere and emits a gamma-ray burst (Section~5, see Figure~\ref{schematic}). The burst radius $r\approx 10^{12}$~cm and Lorentz factor $\Gamma\approx 5$ (the red circle in Figure~1) as well as the predicted burst luminosity $L_\gamma\approx 10^{47}$~erg/s, are in agreement with observations (Section~5.3). Furthermore, the expected average energy of the emitted photons $\bar{E}\sim 10^2$~keV is consistent with observations.

\subsection{Comparison with previous work}

Our model for GRB~170817A shares some features with the shock-breakout models of \cite{Gottlieb2018}, \cite{Bromberg2018}, \cite{Nakar2018b}. However, there are important differences.

(i) The previous models require a small Lorentz factor $\Gamma\sim 1$ in order to explain the observed photon energy $\bar{E}\sim 10^2$~keV. These models adopted the plasma temperature behind the shock $kT\approx 50$~keV, regulated by $e^\pm$ creation, as discussed in earlier papers (e.g. \citealt{Nakar2012}). Then the observed average photon energy $\bar{E}\approx 3kT\,\Gamma\approx 150\,\Gamma$~keV implies $\Gamma\sim 1$. By contrast, we find that the observed light curve requires $\Gamma\simgt 5$. Therefore, we conclude that the shock model with $kT\approx 50$~keV is in tension with observations of GRB~170817A.

(ii) The previous models assumed that shock breakout occurs in a photon-poor cloud, $n_\gamma/n_b\ll10^3$. By contrast, the envelope inflated by the mechanism described in Section~4 is photon-rich, $n_\gamma/n_b>10^4$. Then the delayed explosion in the envelope emits a spectrum with a reduced energy per photon in the jet rest frame, which is consistent with the high-$\Gamma$ Doppler boost giving the observed $\bar{E}\sim 10^2$~keV. We have not calculated yet the detailed GRB spectrum expected from shock breakout in the photon-rich envelope.
A similar problem was studied by \cite{Levinson2012}, and we are currently working on complete, first-principle simulations that will give the emitted spectra with various $n_\gamma/n_b\gg 10^3$. The results will be reported in a future paper. Note that our photon-rich model neglected the photon number generated by the shock itself via downstream bremsstrahlung emission  while the previous models relied on this emission. The self-generation of photons could reach the required $n_\gamma/n_b\sim 10^4$ if the shock is slower and/or the outer layers of the envelope manage to synthesize heavy nuclei \citep{Nakar2019}.

(iii) In the previous models, the shock acceleration and the production of gamma-rays occurred at a well defined characteristic radius --- the cloud ``edge'' where density suddenly and steeply drops by many orders of magnitude. In this respect, the models were similar to the canonical shock breakout in a stellar explosion. By contrast, the envelope described in this paper is equivalent to an infinite medium. The fact that at any given time the envelope extends to a finite radius becomes irrelevant, since its leading edge 
has a diverging Lorentz factor and is out of reach for a blast wave.

The expanding envelope may be idealized as a flow  ejected impulsively from the center with a self-similar (power-law) distribution of Lorentz factor $\gamma(m)$. Its density profile is determined by $\gamma(m)$ and is also self-similar (Section~5.1).  The acceleration of a blast wave launched in such an envelope occurs over a few decades in radius rather than at an edge of a cloud. This qualitatively changes the dynamics and radiation of the blast wave. It has to ``chase'' each layer of the envelope, and catches up with layers of higher $\gamma$ at progressively larger radii $r\propto \gamma^2$ until  finally reaching the photosphere.

\subsection{Future observational tests}

Our results suggest a few observational implications that may be tested in the future. 

(1) Our model for off-axis short GRBs predicts that the relative width of the gamma-ray pulse $\Delta\tobs/\tobs$ reflects the shock strength at the photosphere (\Eq~\ref{eq:width}). The blast wave power is expected to decrease with the viewing polar angle $\theta$. This suggests that the luminosity of the gamma-ray counterpart should decrease with $\theta$ while $\Delta\tobs/\tobs$ should increase. Our model also predicts an anti-correlation between the pulse hardness $\bar{E}$ and relative duration $\Delta\tobs/\tobs$, as both are controlled by the blast wave strength at the photosphere, $\Gph/\gph$. These correlations may be tested by future observations. The  blast wave is fastest when viewed on-axis, directly in front of the powerful collimated jet with $\Gamma\simgt 10^2$. Its gamma-ray emission is also shortest when viewed on-axis, $\Delta\tobs/\tobs\ll 1$. However, the on-axis luminosity from the blast wave should be outshined by the extremely bright, beamed emission from the jet itself, and therefore the above correlations should break at small $\theta$ where the jet comes into view. The jet is expected to emit a canonical short GRB many orders of magnitude brighter than GRB~170817A.

(2) Relativistic ablation of the neutron star surface at the onset of the merger (\Sect~3) suggests an immediate gamma-ray burst, overlapping with the gravitational wave signal. Relativistic ablation creates an ultra-relativistic fireball in a short time $t\sim 3\times 10^{-5}$~s after the two stars touch. Its energy $\En_f$ has a flat distribution over $\gamma$, up to enormous $\gamma\sim 10^3$ (Figure~5). Its outermost, fastest layers become transparent while still being radiation dominated, $w\simgt 1$, and hence radiate away most of their energy, similar to the fireball models of \cite{Paczynski1986} and \cite{Goodman1986}. Thus, a significant fraction of the ablation fireball energy {$\En_f$} is radiated away. The fireball energy is quite uncertain though. It is sensitive to both the neutrino luminosity from the collision sandwich and the geometric parameter $a$ that describes the effective area of the neutrino-sphere near the stellar surface. In the most optimistic simulation with $a=4$~km this energy is $\En_f\sim 3\times 10^{45}$~erg; then the ablation fireball emits a quasi-thermal burst with luminosity up to $10^{51}$~erg/s and duration $\sim 10^{-5}$~s. Its observed temperature is close to the temperature at the base of the outflow, $kT\sim 3-6$~MeV. Even in the optimistic model, this initial, hard ``ablation burst'' is weak and difficult to detect, however it might become detectable with future, more sensitive detectors.

(3) The ejecta acceleration by the internal shock simulated in \Sect~4 has observational consequences for the kilonova emission. Optical/IR emission from GW~170817 can be explained as a superposition of a ``blue'' and a ``red'' component, that came from material with different opacities, with and without synthesized lanthanides, see \cite{Kasen2017}. \cite{Waxman2018} pointed out that the data could also be fitted by a model with a simple (fixed, gray) opacity,  if the emitting material was ejected with a power-law velocity distribution $v\propto m^{-\psi}$ with $\psi\sim 0.6$. For the cloud simulated in \Sect~4, both opacity and velocity may be expected to vary. The early internal shock creates a monotonic four-velocity distribution with a changing slope $\psi=-d\ln(\gamma\beta)/d\ln m$ (Figure~7). In the present paper we focused on the outer layers ($m\ll\Mej$), which 
% have $\gamma\beta>1$, $\psi\approx 1/4$, and 
are relevant for the GRB production. The kilonova is emitted by much deeper layers (the massive part of the cloud, $m$ comparable to $\Mej$).
% where $\psi$ is poorly known, as it depends on the internal density profile of the cloud. 
 Qualitatively, one may expect that the variation of speed and density across the post-shock cloud will lead to the emission of blue and red kilonova components. The faster parts of the cloud will have a lower density, fail to produce lanthanides, and emit the blue component. The slower and denser parts may synthesize the lanthanide material of high opacity and emit the red component. However, it is unclear if our simple spherically symmetric simulations are capable of giving enough mass at low velocities needed for the observed red kilonova. The presence of heavier and slower outflow at large polar angles may need to be invoked to increase red emission and explain the GW~170817 observations. We leave the detailed analysis of this topic for future work.

(4) The ejected envelope is eventually decelerated by an external medium and produces afterglow emission for a broad range of viewing angles. Such deceleration afterglow is in general expected for dynamical ejecta from mergers \citep{Nakar2018a,Hotokezaka2018}, regardless of the presence or absence of a collimated jet. Furthermore, \cite{Nakar2018a} argue that the initial slow rise of the afterglow of GW~170817 comes from ejecta moving toward us, along the line of sight, rather than the collimated jet viewed from the side. At present it is unclear if/when the envelope described in this paper can dominate the observed afterglow. Its calculated stratification $\gamma\beta(m)$ (Figure~7) may be used to develop a detailed afterglow model and check if the decelerating envelope could overshine the off-axis emission from the decelerating jet at late stages of the explosion.

\acknowledgements
We thank Ehud Nakar, the referee of this paper, for his comments. A.M.B. is supported by NSF grant AST-1816484, NASA grant NNX15AE26G, 
a grant from the Simons Foundation \#446228, and the Humboldt Foundation. Y.L. is supported by NSF grant AST-1816484. C.L. was supported by the Swedish National Space Board under grant number Dnr. 107/16.

%################################################################

\appendix

\section{Compressional heating at the collision interface}

The pressure growth in the sandwich between the colliding stars implies a strong compressional heating of their surface layers, as seen from the following consideration. Let $\rho_0$ be the initial, pre-shock density of an old layer in the sandwich, and $P_0$ --- its pressure when it was just crossed by the shock. Pressure in low-density layers is strongly dominated 
by radiation (and $e^\pm$ pairs), so their adiabatic index is $\ad\approx 4/3$. 
As long as neutrino cooling is negligible, the layer compression by
increasing $P$ to a higher density $\rho_0^\prime$ occurs adiabatically,
\beq
  \frac{\rho_0^\prime}{\rho_0}=\left(\frac{P_0^\prime}{P_0}\right)^{1/\ad}
  \approx \left(\frac{P}{P_0}\right)^{1/\ad},
\eeq
where we used pressure balance across the sandwich, $P_0^\prime\approx P$. 
Energy per baryon in the compressed layer grows proportionally to
$P_0^\prime/\rho_0^\prime\propto P^{1-1/\ad}$. 

Instead of energy per baryon it is more convenient to consider dimensionless enthalpy 
per unit mass, $w$. The initial $w_0=(P_0+U_0)/\rho_0 c^2$
is related to $\Eb$ defined in \Eq~(\ref{eq:Eb}) by 
\beq
\label{eq:w_0}
  w_0=\frac{4}{3}\,\frac{\Eb}{c^2}.
\eeq
As the sandwich pressure grows to $P\gg P_0$, the dimensionless enthalpy of the layer 
is amplified as
\beq
\label{eq:w}
  \frac{w_0^\prime}{w_0}=\left(\frac{P}{P_0}\right)^{1-1/\ad}\approx\left(\frac{P}{P_0}\right)^{1/4}
     \approx\left(\frac{\rho}{\rho_0}\right)^{1/4},
\eeq
where $\rho\gg\rho_0$ is the present density of matter just upstream of the shocks. 
Note that $\rho$ and $\rho_0$ are the pre-shock densities of different layers; 
$\rho/\rho_0$ should not be confused with the compression factor of the old layer
due to the increasing pressure, $\rho_0^\prime/\rho_0$. The last 
equality in \Eq~(\ref{eq:w}) may be slightly changed (by a numerical factor close to unity)
when the shock pressure becomes dominated by nucleons, which leads to the postshock 
adiabatic index $\ad\approx 5/3$ instead of $4/3$. However, the scaling 
$w_0^\prime/w_0\propto \rho^{1/4}$ is weakly affected by this transition, because the 
scaling applies to the old, low-density layers, which remain radiation-dominated with 
$\ad=4/3$.
 
The moderate initial $w_0<0.1$ increases with time to $w_0^\prime>1$ 
when the shocks propagate into layers of density $\rho>w_0^{-4}\rho_0$. 
The layers compressed to relativistic enthalpy $w_0^\prime\gg 1$ have the potential of being 
ejected with highly relativistic speeds, if their internal energy has a chance to convert 
to bulk kinetic energy via adiabatic expansion without losing it to the neighboring 
heavy and non-relativistic layers.

The maximum compression factor may be reached 
close to the moment when the squeezed matter begins to leak out from the sandwich. 
At the beginning of the collision, the size of the collision interface in the $y$-$z$ plane
expands with a superluminal speed $dr_A/dt$. Here $r_A=(y_A^2+z_A^2)^{1/2}$
represents the curve where the surfaces of the two stars intersect, which defines 
the edge of the sandwich (this curve is not a circle, as the tangential motion of the colliding stars breaks the axial symmetry of the interface). 
The interface area grows with time because of the converging motion of the two stars, 
which brings it into contact with more material. The two shocks bounding the collision sandwich
intersect at $r_A$, which initially grows with rate $dr_A/dt>c$.
Later $dr_A/dt$ is reduced (because of the curvature of the stellar surfaces) and 
eventually becomes subliminal and subsonic. Then the hot interface material pushes 
the shocks aside (so that they no longer intersect at the edge) and leaks out with the 
sound speed through the edges of the sandwich. This mass loss will buffer 
the growth of pressure in the sandwich.

The time at which the sandwich matter begins to be ejected from the edges may be roughly estimated as 
$t\sim R/c\approx 3\times 10^{-5}$~s. During this time, the two shocks propagate a 
significant distance $x\sim t\vsh\simgt 1$~km into the deep stellar interior, where the 
upstream density can be quite large, $\rho\sim 10^{13}-10^{14}$~g~cm$^{-3}$. Therefore, we estimate the maximum possible compression in the sandwich using 
the ram pressure that corresponds to the upstream $\rho\sim 10^{14}$~g~cm$^{-3}$.
The corresponding $\rho_0$ of layers reaching $w_0^\prime>1$ is found from 
\Eq~(\ref{eq:w}), 
\beq
  \rho_0\approx 10^{10}\,\left(\frac{w_0}{0.1}\right)^{-4} \rho_{14} {\rm~g~cm}^{-3}.
\eeq
Recalling that $\rho_0$ is the pre-shock density of sub-surface layers with a hydrostatic scale-height $h\sim 0.3-1$~km,
one can roughly estimate their mass as $m\sim r^2 h\rho_0$, where $r\simlt R\sim 10$~km
is the characteristic extension of the old layers along the interface. The initial volume 
occupied by these layers is $V_0\sim r^2 h\simlt 10^{16}$~cm$^{-3}$, and their nucleon 
mass is
\beq
\label{eq:m_paste}
  m\sim V_0\rho_0\sim 10^{26}\left(\frac{w_0}{0.1}\right)^4 \rho_{14} {\rm ~g}.
\eeq

This mass has the potential of being ejected with a highly relativistic speed. However, a more detailed analysis suggests that this mechanism 
is hindered by two processes in the sandwich between the colliding stars:
(1) The idealized picture where mass $m$ estimated in \Eq~(\ref{eq:m_paste}) resides in a very thin (strongly compressed) layer is likely incorrect, because
the sandwich is also the site of Kelvin-Helmholtz instability. The 
high-$w$ material is likely to be dispersed into small bubbles or filaments and mixed into 
dense, massive, low-$w$ material before escaping the sandwich. 
Only a small fraction of the bubbles might be able to escape with a 
highly relativistic momentum and avoid sharing it with the non-relativistic matter. (2) Neutrino emission and transport tends to steal energy from the layers with high $w$ and reduce their pressure. This amplifies their compression and limits enthalpy per unit rest mass, $w$.

\bibliography{merger}

\begin{thebibliography}{}
\expandafter\ifx\csname natexlab\endcsname\relax\def\natexlab#1{#1}\fi

\bibitem[{{Abbott} {et~al.}(2017{\natexlab{a}}){Abbott}, {Abbott}, {Abbott},
  {Acernese}, {Ackley}, {Adams}, {Adams}, {Addesso}, {Adhikari}, {Adya}, \&
  et~al.}]{Abbott2017a}
{Abbott}, B.~P., {Abbott}, R., {Abbott}, T.~D., {et~al.} 2017{\natexlab{a}},
  \apjl, 848, L13

\bibitem[{{Abbott} {et~al.}(2017{\natexlab{b}}){Abbott}, {Abbott}, {Abbott},
  {Acernese}, {Ackley}, {Adams}, {Adams}, {Addesso}, {Adhikari}, {Adya}, \&
  et~al.}]{Abbott2017b}
---. 2017{\natexlab{b}}, Physical Review Letters, 119, 161101

\bibitem[{{Abramowicz} {et~al.}(1991){Abramowicz}, {Novikov}, \&
  {Paczynski}}]{Abramowicz1991}
{Abramowicz}, M.~A., {Novikov}, I.~D., \& {Paczynski}, B. 1991, \apj, 369, 175

\bibitem[{{Bauswein} {et~al.}(2013){Bauswein}, {Goriely}, \&
  {Janka}}]{Bauswein2013}
{Bauswein}, A., {Goriely}, S., \& {Janka}, H.-T. 2013, \apj, 773, 78

\bibitem[{{Beloborodov}(2003)}]{Beloborodov2003}
{Beloborodov}, A.~M. 2003, \apj, 588, 931

\bibitem[{{Beloborodov}(2010)}]{Beloborodov2010}
---. 2010, \mnras, 407, 1033

\bibitem[{{Beloborodov}(2011)}]{Beloborodov2011}
---. 2011, \apj, 737, 68

\bibitem[{{Beloborodov}(2013)}]{Beloborodov2013}
---. 2013, \apj, 764, 157

\bibitem[{{Beloborodov}(2017)}]{Beloborodov2017}
---. 2017, \apj, 838, 125

\bibitem[{{Belyanin} {et~al.}(2001){Belyanin}, {Derishev}, {Kocharovsky}, \&
  {Kocharovsky}}]{Belyanin2001}
{Belyanin}, A.~A., {Derishev}, E.~V., {Kocharovsky}, V.~V., \& {Kocharovsky},
  V.~V. 2001, arXiv Astrophysics e-prints, astro-ph/0111558

\bibitem[{{Birkl} {et~al.}(2007){Birkl}, {Aloy}, {Janka}, \&
  {M{\"u}ller}}]{Birkl2007}
{Birkl}, R., {Aloy}, M.~A., {Janka}, H.-T., \& {M{\"u}ller}, E. 2007, \aap,
  463, 51

\bibitem[{{Bromberg} {et~al.}(2018){Bromberg}, {Tchekhovskoy}, {Gottlieb},
  {Nakar}, \& {Piran}}]{Bromberg2018}
{Bromberg}, O., {Tchekhovskoy}, A., {Gottlieb}, O., {Nakar}, E., \& {Piran}, T.
  2018, \mnras, 475, 2971

\bibitem[{{Chen} \& {Beloborodov}(2007)}]{Chen_Beloborodov2007}
{Chen}, W.-X., \& {Beloborodov}, A.~M. 2007, \apj, 657, 383

\bibitem[{{Coulter} {et~al.}(2017){Coulter}, {Foley}, {Kilpatrick}, {Drout},
  {Piro}, {Shappee}, {Siebert}, {Simon}, {Ulloa}, {Kasen}, {Madore},
  {Murguia-Berthier}, {Pan}, {Prochaska}, {Ramirez-Ruiz}, {Rest}, \&
  {Rojas-Bravo}}]{Coulter2017}
{Coulter}, D.~A., {Foley}, R.~J., {Kilpatrick}, C.~D., {et~al.} 2017, Science,
  358, 1556

\bibitem[{{Dessart} {et~al.}(2009){Dessart}, {Ott}, {Burrows}, {Rosswog}, \&
  {Livne}}]{Dessart2009}
{Dessart}, L., {Ott}, C.~D., {Burrows}, A., {Rosswog}, S., \& {Livne}, E. 2009,
  \apj, 690, 1681

\bibitem[{{Drout} {et~al.}(2017){Drout}, {Piro}, {Shappee}, {Kilpatrick},
  {Simon}, {Contreras}, {Coulter}, {Foley}, {Siebert}, {Morrell}, {Boutsia},
  {Di Mille}, {Holoien}, {Kasen}, {Kollmeier}, {Madore}, {Monson},
  {Murguia-Berthier}, {Pan}, {Prochaska}, {Ramirez-Ruiz}, {Rest}, {Adams},
  {Alatalo}, {Ba{\~n}ados}, {Baughman}, {Beers}, {Bernstein}, {Bitsakis},
  {Campillay}, {Hansen}, {Higgs}, {Ji}, {Maravelias}, {Marshall}, {Moni Bidin},
  {Prieto}, {Rasmussen}, {Rojas-Bravo}, {Strom}, {Ulloa},
  {Vargas-Gonz{\'a}lez}, {Wan}, \& {Whitten}}]{Drout2017}
{Drout}, M.~R., {Piro}, A.~L., {Shappee}, B.~J., {et~al.} 2017, Science, 358,
  1570

\bibitem[{{Duffell} {et~al.}(2018){Duffell}, {Quataert}, {Kasen}, \&
  {Klion}}]{Duffell2018}
{Duffell}, P.~C., {Quataert}, E., {Kasen}, D., \& {Klion}, H. 2018, \apj, 866,
  3

\bibitem[{{Evans} {et~al.}(2017){Evans}, {Cenko}, {Kennea}, {Emery}, {Kuin},
  {Korobkin}, {Wollaeger}, {Fryer}, {Madsen}, {Harrison}, {Xu}, {Nakar},
  {Hotokezaka}, {Lien}, {Campana}, {Oates}, {Troja}, {Breeveld}, {Marshall},
  {Barthelmy}, {Beardmore}, {Burrows}, {Cusumano}, {D'A{\`i}}, {D'Avanzo},
  {D'Elia}, {de Pasquale}, {Even}, {Fontes}, {Forster}, {Garcia}, {Giommi},
  {Grefenstette}, {Gronwall}, {Hartmann}, {Heida}, {Hungerford}, {Kasliwal},
  {Krimm}, {Levan}, {Malesani}, {Melandri}, {Miyasaka}, {Nousek}, {O'Brien},
  {Osborne}, {Pagani}, {Page}, {Palmer}, {Perri}, {Pike}, {Racusin}, {Rosswog},
  {Siegel}, {Sakamoto}, {Sbarufatti}, {Tagliaferri}, {Tanvir}, \&
  {Tohuvavohu}}]{Evans2017}
{Evans}, P.~A., {Cenko}, S.~B., {Kennea}, J.~A., {et~al.} 2017, Science, 358,
  1565

\bibitem[{{Goldstein} {et~al.}(2017){Goldstein}, {Veres}, {Burns}, {Briggs},
  {Hamburg}, {Kocevski}, {Wilson-Hodge}, {Preece}, {Poolakkil}, {Roberts},
  {Hui}, {Connaughton}, {Racusin}, {von Kienlin}, {Dal Canton}, {Christensen},
  {Littenberg}, {Siellez}, {Blackburn}, {Broida}, {Bissaldi}, {Cleveland},
  {Gibby}, {Giles}, {Kippen}, {McBreen}, {McEnery}, {Meegan}, {Paciesas}, \&
  {Stanbro}}]{Goldstein2017}
{Goldstein}, A., {Veres}, P., {Burns}, E., {et~al.} 2017, \apjl, 848, L14

\bibitem[{{Goodman}(1986)}]{Goodman1986}
{Goodman}, J. 1986, \apjl, 308, L47

\bibitem[{{Goodman} {et~al.}(1987){Goodman}, {Dar}, \&
  {Nussinov}}]{Goodman1987}
{Goodman}, J., {Dar}, A., \& {Nussinov}, S. 1987, \apjl, 314, L7

\bibitem[{{Gottlieb} {et~al.}(2018){Gottlieb}, {Nakar}, {Piran}, \&
  {Hotokezaka}}]{Gottlieb2018}
{Gottlieb}, O., {Nakar}, E., {Piran}, T., \& {Hotokezaka}, K. 2018, \mnras,
  479, 588

\bibitem[{{Granot} {et~al.}(2018){Granot}, {Gill}, {Guetta}, \& {De
  Colle}}]{Granot2018}
{Granot}, J., {Gill}, R., {Guetta}, D., \& {De Colle}, F. 2018, \mnras, 481,
  1597

\bibitem[{{Hallinan} {et~al.}(2017){Hallinan}, {Corsi}, {Mooley}, {Hotokezaka},
  {Nakar}, {Kasliwal}, {Kaplan}, {Frail}, {Myers}, {Murphy}, {De}, {Dobie},
  {Allison}, {Bannister}, {Bhalerao}, {Chandra}, {Clarke}, {Giacintucci}, {Ho},
  {Horesh}, {Kassim}, {Kulkarni}, {Lenc}, {Lockman}, {Lynch}, {Nichols},
  {Nissanke}, {Palliyaguru}, {Peters}, {Piran}, {Rana}, {Sadler}, \&
  {Singer}}]{Hallinan2017}
{Hallinan}, G., {Corsi}, A., {Mooley}, K.~P., {et~al.} 2017, Science, 358, 1579

\bibitem[{{Hotokezaka} {et~al.}(2013){Hotokezaka}, {Kiuchi}, {Kyutoku},
  {Okawa}, {Sekiguchi}, {Shibata}, \& {Taniguchi}}]{Hotokezaka2013}
{Hotokezaka}, K., {Kiuchi}, K., {Kyutoku}, K., {et~al.} 2013, \prd, 87, 024001

\bibitem[{{Hotokezaka} {et~al.}(2018){Hotokezaka}, {Kiuchi}, {Shibata},
  {Nakar}, \& {Piran}}]{Hotokezaka2018}
{Hotokezaka}, K., {Kiuchi}, K., {Shibata}, M., {Nakar}, E., \& {Piran}, T.
  2018, \apj, 867, 95

\bibitem[{{Ito} {et~al.}(2018){Ito}, {Levinson}, {Stern}, \&
  {Nagataki}}]{Ito2018}
{Ito}, H., {Levinson}, A., {Stern}, B.~E., \& {Nagataki}, S. 2018, \mnras, 474,
  2828

\bibitem[{{Johnson} \& {McKee}(1971)}]{Johnson1971}
{Johnson}, M.~H., \& {McKee}, C.~F. 1971, \prd, 3, 858

\bibitem[{{Kasen} {et~al.}(2017){Kasen}, {Metzger}, {Barnes}, {Quataert}, \&
  {Ramirez-Ruiz}}]{Kasen2017}
{Kasen}, D., {Metzger}, B., {Barnes}, J., {Quataert}, E., \& {Ramirez-Ruiz}, E.
  2017, \nat, 551, 80

\bibitem[{{Kasliwal} {et~al.}(2017){Kasliwal}, {Nakar}, {Singer}, {Kaplan},
  {Cook}, {Van Sistine}, {Lau}, {Fremling}, {Gottlieb}, {Jencson}, {Adams},
  {Feindt}, {Hotokezaka}, {Ghosh}, {Perley}, {Yu}, {Piran}, {Allison},
  {Anupama}, {Balasubramanian}, {Bannister}, {Bally}, {Barnes}, {Barway},
  {Bellm}, {Bhalerao}, {Bhattacharya}, {Blagorodnova}, {Bloom}, {Brady},
  {Cannella}, {Chatterjee}, {Cenko}, {Cobb}, {Copperwheat}, {Corsi}, {De},
  {Dobie}, {Emery}, {Evans}, {Fox}, {Frail}, {Frohmaier}, {Goobar}, {Hallinan},
  {Harrison}, {Helou}, {Hinderer}, {Ho}, {Horesh}, {Ip}, {Itoh}, {Kasen},
  {Kim}, {Kuin}, {Kupfer}, {Lynch}, {Madsen}, {Mazzali}, {Miller}, {Mooley},
  {Murphy}, {Ngeow}, {Nichols}, {Nissanke}, {Nugent}, {Ofek}, {Qi}, {Quimby},
  {Rosswog}, {Rusu}, {Sadler}, {Schmidt}, {Sollerman}, {Steele}, {Williamson},
  {Xu}, {Yan}, {Yatsu}, {Zhang}, \& {Zhao}}]{Kasliwal2017}
{Kasliwal}, M.~M., {Nakar}, E., {Singer}, L.~P., {et~al.} 2017, Science, 358,
  1559

\bibitem[{{Kiuchi} {et~al.}(2018){Kiuchi}, {Kyutoku}, {Sekiguchi}, \&
  {Shibata}}]{Kiuchi2018}
{Kiuchi}, K., {Kyutoku}, K., {Sekiguchi}, Y., \& {Shibata}, M. 2018, \prd, 97,
  124039

\bibitem[{{Klu{\'z}niak} \& {Ruderman}(1998)}]{Kluzniak1998}
{Klu{\'z}niak}, W., \& {Ruderman}, M. 1998, \apjl, 505, L113

\bibitem[{{Kyutoku} {et~al.}(2014){Kyutoku}, {Ioka}, \&
  {Shibata}}]{Kyutoku2014}
{Kyutoku}, K., {Ioka}, K., \& {Shibata}, M. 2014, \mnras, 437, L6

\bibitem[{{Lamb} {et~al.}(2018){Lamb}, {Mandel}, \& {Resmi}}]{Lamb2018}
{Lamb}, G.~P., {Mandel}, I., \& {Resmi}, L. 2018, \mnras, 481, 2581

\bibitem[{{Lazzati} {et~al.}(2017{\natexlab{a}}){Lazzati}, {Deich}, {Morsony},
  \& {Workman}}]{Lazzati2017a}
{Lazzati}, D., {Deich}, A., {Morsony}, B.~J., \& {Workman}, J.~C.
  2017{\natexlab{a}}, \mnras, 471, 1652

\bibitem[{{Lazzati} {et~al.}(2017{\natexlab{b}}){Lazzati},
  {L{\'o}pez-C{\'a}mara}, {Cantiello}, {Morsony}, {Perna}, \&
  {Workman}}]{Lazzati2017b}
{Lazzati}, D., {L{\'o}pez-C{\'a}mara}, D., {Cantiello}, M., {et~al.}
  2017{\natexlab{b}}, \apjl, 848, L6

\bibitem[{{Lazzati} {et~al.}(2018){Lazzati}, {Perna}, {Morsony},
  {Lopez-Camara}, {Cantiello}, {Ciolfi}, {Giacomazzo}, \&
  {Workman}}]{Lazzati2018}
{Lazzati}, D., {Perna}, R., {Morsony}, B.~J., {et~al.} 2018, Physical Review
  Letters, 120, 241103

\bibitem[{{Levinson}(2012)}]{Levinson2012}
{Levinson}, A. 2012, \apj, 756, 174

\bibitem[{{Lipunova} \& {Lipunov}(1998)}]{Lipunova1998}
{Lipunova}, G.~V., \& {Lipunov}, V.~M. 1998, \aap, 329, L29

\bibitem[{{Lithwick} \& {Sari}(2001)}]{Lithwick2001}
{Lithwick}, Y., \& {Sari}, R. 2001, \apj, 555, 540

\bibitem[{{Lundman} {et~al.}(2018){Lundman}, {Beloborodov}, \&
  {Vurm}}]{Lundman2018}
{Lundman}, C., {Beloborodov}, A.~M., \& {Vurm}, I. 2018, \apj, 858, 7

\bibitem[{{Margutti} {et~al.}(2017){Margutti}, {Berger}, {Fong}, {Guidorzi},
  {Alexander}, {Metzger}, {Blanchard}, {Cowperthwaite}, {Chornock},
  {Eftekhari}, {Nicholl}, {Villar}, {Williams}, {Annis}, {Brown}, {Chen},
  {Doctor}, {Frieman}, {Holz}, {Sako}, \& {Soares-Santos}}]{Margutti2017}
{Margutti}, R., {Berger}, E., {Fong}, W., {et~al.} 2017, \apjl, 848, L20

\bibitem[{{Matsumoto} {et~al.}(2019){Matsumoto}, {Nakar}, \&
  {Piran}}]{Matsumoto2019}
{Matsumoto}, T., {Nakar}, E., \& {Piran}, T. 2019, \mnras, 483, 1247

\bibitem[{{Matzner} {et~al.}(2013){Matzner}, {Levin}, \& {Ro}}]{Matzner2013}
{Matzner}, C.~D., {Levin}, Y., \& {Ro}, S. 2013, \apj, 779, 60

\bibitem[{{Mestel} \& {Spruit}(1987)}]{Mestel1987}
{Mestel}, L., \& {Spruit}, H.~C. 1987, \mnras, 226, 57

\bibitem[{{Metzger} {et~al.}(2007){Metzger}, {Thompson}, \&
  {Quataert}}]{Metzger2007}
{Metzger}, B.~D., {Thompson}, T.~A., \& {Quataert}, E. 2007, \apj, 659, 561

\bibitem[{{Metzger} {et~al.}(2018){Metzger}, {Thompson}, \&
  {Quataert}}]{Metzger2018}
---. 2018, \apj, 856, 101

\bibitem[{{Metzger} {et~al.}(2010){Metzger}, {Mart{\'{\i}}nez-Pinedo},
  {Darbha}, {Quataert}, {Arcones}, {Kasen}, {Thomas}, {Nugent}, {Panov}, \&
  {Zinner}}]{Metzger2010}
{Metzger}, B.~D., {Mart{\'{\i}}nez-Pinedo}, G., {Darbha}, S., {et~al.} 2010,
  \mnras, 406, 2650

\bibitem[{{Mooley} {et~al.}(2018{\natexlab{a}}){Mooley}, {Frail}, {Dobie},
  {Lenc}, {Corsi}, {De}, {Nayana}, {Makhathini}, {Heywood}, {Murphy}, {Kaplan},
  {Chandra}, {Smirnov}, {Nakar}, {Hallinan}, {Camilo}, {Fender}, {Goedhart},
  {Groot}, {Kasliwal}, {Kulkarni}, \& {Woudt}}]{Mooley2018b}
{Mooley}, K.~P., {Frail}, D.~A., {Dobie}, D., {et~al.} 2018{\natexlab{a}},
  \apjl, 868, L11

\bibitem[{{Mooley} {et~al.}(2018{\natexlab{b}}){Mooley}, {Deller}, {Gottlieb},
  {Nakar}, {Hallinan}, {Bourke}, {Frail}, {Horesh}, {Corsi}, \&
  {Hotokezaka}}]{Mooley2018a}
{Mooley}, K.~P., {Deller}, A.~T., {Gottlieb}, O., {et~al.} 2018{\natexlab{b}},
  \nat, 561, 355

\bibitem[{{Nakar}(2007)}]{Nakar2007}
{Nakar}, E. 2007, \physrep, 442, 166

\bibitem[{{Nakar}(2019)}]{Nakar2019}
---. 2019, arXiv e-prints, arXiv:1912.05659

\bibitem[{{Nakar} {et~al.}(2018){Nakar}, {Gottlieb}, {Piran}, {Kasliwal}, \&
  {Hallinan}}]{Nakar2018b}
{Nakar}, E., {Gottlieb}, O., {Piran}, T., {Kasliwal}, M.~M., \& {Hallinan}, G.
  2018, \apj, 867, 18

\bibitem[{{Nakar} \& {Piran}(2018)}]{Nakar2018a}
{Nakar}, E., \& {Piran}, T. 2018, \mnras, 478, 407

\bibitem[{{Nakar} \& {Sari}(2012)}]{Nakar2012}
{Nakar}, E., \& {Sari}, R. 2012, \apj, 747, 88

\bibitem[{{Paczynski}(1986)}]{Paczynski1986}
{Paczynski}, B. 1986, \apjl, 308, L43

\bibitem[{{Paczy{\'n}ski}(1998)}]{Paczynski1998}
{Paczy{\'n}ski}, B. 1998, in American Institute of Physics Conference Series,
  Vol. 428, Gamma-Ray Bursts, 4th Hunstville Symposium, ed. C.~A. {Meegan},
  R.~D. {Preece}, \& T.~M. {Koshut}, 783--787

\bibitem[{{Pan} \& {Sari}(2006)}]{Pan2006}
{Pan}, M., \& {Sari}, R. 2006, \apj, 643, 416

\bibitem[{{Pozanenko} {et~al.}(2018){Pozanenko}, {Barkov}, {Minaev}, {Volnova},
  {Mazaeva}, {Moskvitin}, {Krugov}, {Samodurov}, {Loznikov}, \&
  {Lyutikov}}]{Pozanenko2018}
{Pozanenko}, A.~S., {Barkov}, M.~V., {Minaev}, P.~Y., {et~al.} 2018, \apjl,
  852, L30

\bibitem[{{Price} \& {Rosswog}(2006)}]{Price2006}
{Price}, D.~J., \& {Rosswog}, S. 2006, Science, 312, 719

\bibitem[{{Qian} \& {Woosley}(1996)}]{Qian1996}
{Qian}, Y.-Z., \& {Woosley}, S.~E. 1996, \apj, 471, 331

\bibitem[{{Radice} {et~al.}(2016){Radice}, {Galeazzi}, {Lippuner}, {Roberts},
  {Ott}, \& {Rezzolla}}]{Radice2016}
{Radice}, D., {Galeazzi}, F., {Lippuner}, J., {et~al.} 2016, \mnras, 460, 3255

\bibitem[{{Radice} {et~al.}(2018){Radice}, {Perego}, {Hotokezaka}, {Fromm},
  {Bernuzzi}, \& {Roberts}}]{Radice2018}
{Radice}, D., {Perego}, A., {Hotokezaka}, K., {et~al.} 2018, arXiv e-prints,
  arXiv:1809.11161

\bibitem[{{Savchenko} {et~al.}(2017){Savchenko}, {Ferrigno}, {Kuulkers},
  {Bazzano}, {Bozzo}, {Brandt}, {Chenevez}, {Courvoisier}, {Diehl}, {Domingo},
  {Hanlon}, {Jourdain}, {von Kienlin}, {Laurent}, {Lebrun}, {Lutovinov},
  {Martin-Carrillo}, {Mereghetti}, {Natalucci}, {Rodi}, {Roques}, {Sunyaev}, \&
  {Ubertini}}]{Savchenko2017}
{Savchenko}, V., {Ferrigno}, C., {Kuulkers}, E., {et~al.} 2017, \apjl, 848, L15

\bibitem[{{Siegel} \& {Metzger}(2017)}]{Siegel2017}
{Siegel}, D.~M., \& {Metzger}, B.~D. 2017, Physical Review Letters, 119, 231102

\bibitem[{{Soares-Santos} {et~al.}(2017){Soares-Santos}, {Holz}, {Annis},
  {Chornock}, {Herner}, {Berger}, {Brout}, {Chen}, {Kessler}, {Sako}, {Allam},
  {Tucker}, {Butler}, {Palmese}, {Doctor}, {Diehl}, {Frieman}, {Yanny}, {Lin},
  {Scolnic}, {Cowperthwaite}, {Neilsen}, {Marriner}, {Kuropatkin}, {Hartley},
  {Paz-Chinch{\'o}n}, {Alexander}, {Balbinot}, {Blanchard}, {Brown}, {Carlin},
  {Conselice}, {Cook}, {Drlica-Wagner}, {Drout}, {Durret}, {Eftekhari}, {Farr},
  {Finley}, {Foley}, {Fong}, {Fryer}, {Garc{\'{\i}}a-Bellido}, {Gill},
  {Gruendl}, {Hanna}, {Kasen}, {Li}, {Lopes}, {Louren{\c c}o}, {Margutti},
  {Marshall}, {Matheson}, {Medina}, {Metzger}, {Mu{\~n}oz}, {Muir}, {Nicholl},
  {Quataert}, {Rest}, {Sauseda}, {Schlegel}, {Secco}, {Sobreira}, {Stebbins},
  {Villar}, {Vivas}, {Walker}, {Wester}, {Williams}, {Zenteno}, {Zhang},
  {Abbott}, {Abdalla}, {Banerji}, {Bechtol}, {Benoit-L{\'e}vy}, {Bertin},
  {Brooks}, {Buckley-Geer}, {Burke}, {Carnero Rosell}, {Carrasco Kind},
  {Carretero}, {Castander}, {Crocce}, {Cunha}, {D'Andrea}, {da Costa}, {Davis},
  {Desai}, {Dietrich}, {Doel}, {Eifler}, {Fernandez}, {Flaugher}, {Fosalba},
  {Gaztanaga}, {Gerdes}, {Giannantonio}, {Goldstein}, {Gruen}, {Gschwend},
  {Gutierrez}, {Honscheid}, {Jain}, {James}, {Jeltema}, {Johnson}, {Johnson},
  {Kent}, {Krause}, {Kron}, {Kuehn}, {Kuhlmann}, {Lahav}, {Lima}, {Maia},
  {March}, {McMahon}, {Menanteau}, {Miquel}, {Mohr}, {Nichol}, {Nord},
  {Ogando}, {Petravick}, {Plazas}, {Romer}, {Roodman}, {Rykoff}, {Sanchez},
  {Scarpine}, {Schubnell}, {Sevilla-Noarbe}, {Smith}, {Smith}, {Suchyta},
  {Swanson}, {Tarle}, {Thomas}, {Thomas}, {Troxel}, {Vikram}, {Wechsler},
  {Weller}, {Dark Energy Survey}, \& {Dark Energy Camera GW-EM
  Collaboration}}]{Soares-Santos2017}
{Soares-Santos}, M., {Holz}, D.~E., {Annis}, J., {et~al.} 2017, \apjl, 848, L16

\bibitem[{{Tan} {et~al.}(2001){Tan}, {Matzner}, \& {McKee}}]{Tan2001}
{Tan}, J.~C., {Matzner}, C.~D., \& {McKee}, C.~F. 2001, \apj, 551, 946

\bibitem[{{Tanaka} {et~al.}(2017){Tanaka}, {Utsumi}, {Mazzali}, {Tominaga},
  {Yoshida}, {Sekiguchi}, {Morokuma}, {Motohara}, {Ohta}, {Kawabata}, {Abe},
  {Aoki}, {Asakura}, {Baar}, {Barway}, {Bond}, {Doi}, {Fujiyoshi}, {Furusawa},
  {Honda}, {Itoh}, {Kawabata}, {Kawai}, {Kim}, {Lee}, {Miyazaki}, {Morihana},
  {Nagashima}, {Nagayama}, {Nakaoka}, {Nakata}, {Ohsawa}, {Ohshima}, {Okita},
  {Saito}, {Sumi}, {Tajitsu}, {Takahashi}, {Takayama}, {Tamura}, {Tanaka},
  {Terai}, {Tristram}, {Yasuda}, \& {Zenko}}]{Tanaka2017}
{Tanaka}, M., {Utsumi}, Y., {Mazzali}, P.~A., {et~al.} 2017, \pasj, 69, 102

\bibitem[{{Tanvir} {et~al.}(2017){Tanvir}, {Levan},
  {Gonz{\'a}lez-Fern{\'a}ndez}, {Korobkin}, {Mandel}, {Rosswog}, {Hjorth},
  {D'Avanzo}, {Fruchter}, {Fryer}, {Kangas}, {Milvang-Jensen}, {Rosetti},
  {Steeghs}, {Wollaeger}, {Cano}, {Copperwheat}, {Covino}, {D'Elia}, {de Ugarte
  Postigo}, {Evans}, {Even}, {Fairhurst}, {Figuera Jaimes}, {Fontes}, {Fujii},
  {Fynbo}, {Gompertz}, {Greiner}, {Hodosan}, {Irwin}, {Jakobsson},
  {J{\o}rgensen}, {Kann}, {Lyman}, {Malesani}, {McMahon}, {Melandri},
  {O'Brien}, {Osborne}, {Palazzi}, {Perley}, {Pian}, {Piranomonte}, {Rabus},
  {Rol}, {Rowlinson}, {Schulze}, {Sutton}, {Th{\"o}ne}, {Ulaczyk}, {Watson},
  {Wiersema}, \& {Wijers}}]{Tanvir2017}
{Tanvir}, N.~R., {Levan}, A.~J., {Gonz{\'a}lez-Fern{\'a}ndez}, C., {et~al.}
  2017, \apjl, 848, L27

\bibitem[{{Thompson} {et~al.}(2001){Thompson}, {Burrows}, \&
  {Meyer}}]{Thompson2001}
{Thompson}, T.~A., {Burrows}, A., \& {Meyer}, B.~S. 2001, \apj, 562, 887

\bibitem[{{Thompson} {et~al.}(2004){Thompson}, {Chang}, \&
  {Quataert}}]{Thompson2004}
{Thompson}, T.~A., {Chang}, P., \& {Quataert}, E. 2004, \apj, 611, 380

\bibitem[{{Troja} {et~al.}(2017){Troja}, {Piro}, {van Eerten}, {Wollaeger},
  {Im}, {Fox}, {Butler}, {Cenko}, {Sakamoto}, {Fryer}, {Ricci}, {Lien}, {Ryan},
  {Korobkin}, {Lee}, {Burgess}, {Lee}, {Watson}, {Choi}, {Covino}, {D'Avanzo},
  {Fontes}, {Gonz{\'a}lez}, {Khandrika}, {Kim}, {Kim}, {Lee}, {Lee}, {Kutyrev},
  {Lim}, {S{\'a}nchez-Ram{\'{\i}}rez}, {Veilleux}, {Wieringa}, \&
  {Yoon}}]{Troja2017}
{Troja}, E., {Piro}, L., {van Eerten}, H., {et~al.} 2017, \nat, 551, 71

\bibitem[{{Veres} {et~al.}(2018){Veres}, {M{\'e}sz{\'a}ros}, {Goldstein},
  {Fraija}, {Connaughton}, {Burns}, {Preece}, {Hamburg}, {Wilson-Hodge},
  {Briggs}, \& {Kocevski}}]{Veres2018}
{Veres}, P., {M{\'e}sz{\'a}ros}, P., {Goldstein}, A., {et~al.} 2018, arXiv
  e-prints, arXiv:1802.07328

\bibitem[{{Vurm} \& {Beloborodov}(2016)}]{Vurm2016}
{Vurm}, I., \& {Beloborodov}, A.~M. 2016, \apj, 831, 175

\bibitem[{{Waxman} {et~al.}(2018){Waxman}, {Ofek}, {Kushnir}, \&
  {Gal-Yam}}]{Waxman2018}
{Waxman}, E., {Ofek}, E.~O., {Kushnir}, D., \& {Gal-Yam}, A. 2018, \mnras, 481,
  3423

\bibitem[{{Xie} {et~al.}(2018){Xie}, {Zrake}, \& {MacFadyen}}]{Xie2018}
{Xie}, X., {Zrake}, J., \& {MacFadyen}, A. 2018, \apj, 863, 58

\bibitem[{{Zalamea} \& {Beloborodov}(2011)}]{Zalamea2011}
{Zalamea}, I., \& {Beloborodov}, A.~M. 2011, \mnras, 410, 2302

\bibitem[{{Zel'dovich} \& {Raizer}(1967)}]{Zeldovich1967}
{Zel'dovich}, Y.~B., \& {Raizer}, Y.~P. 1967, {Physics of shock waves and
  high-temperature hydrodynamic phenomena}

\bibitem[{{Zrake} \& {MacFadyen}(2013)}]{Zrake2013}
{Zrake}, J., \& {MacFadyen}, A.~I. 2013, \apjl, 769, L29

\end{thebibliography}

 \end{document}